\documentstyle[epsfig,amssymb]{article}
\begin{document}

\title{Geometry, stochastic calculus and quantum fields in a non-commutative
space-time}
\author{R. Vilela Mendes \\
{\small Grupo de F\'{\i }sica Matem\'{a}tica}\\
{\small \ Complexo Interdisciplinar, Universidade de Lisboa, }\\
{\small \ Av. Gama Pinto, 2 - P1699 Lisboa Codex, Portugal}}
\date{}
\maketitle

\begin{abstract}
The algebras of non-relativistic and of classical mechanics are unstable
algebraic structures. Their deformation towards stable structures leads,
respectively, to relativity and to quantum mechanics. Likewise, the combined
relativistic quantum mechanics algebra is also unstable. Its stabilization
requires the non-commutativity of the space-time coordinates and the
existence of a fundamental length constant.

The new relativistic quantum mechanics algebra has important consequences on
the geometry of space-time, on quantum stochastic calculus and on the
construction of quantum fields. Some of these effects are studied in this
paper.
\end{abstract}

\section{The instability of relativistic quantum mechanics and a fundamental
length}

Physical models and theories are mere approximations to Nature and the
physical constants can never be known with absolute precision. Therefore, if
a fine tuning of the parameters is needed to reproduce some particular
phenomenon, it is probable that the model is basically unsound and that its
other predictions are unreliable. A wider range of validity is expected for
theories that do not change in a qualitative manner for a small change of
parameters. Such theories are called {\it stable} or {\it rigid}.

A mathematical structure is said to be {\it stable} (or {\it rigid}) for a
class of {\it deformations} if any deformation in this class leads to an
equivalent (isomorphic) structure. The idea of stability of structures
provides a guiding principle to test either the validity or the need for
generalization of a physical theory. Namely, if the mathematical structure
of a given theory turns out to be unstable, one might just as well deform
it, until one falls into a stable one, which has a good chance of being a
generalization of wider validity.

The stable-model point of view had a large impact in the field of non-linear
dynamics, where it led to the notion of {\it structural stability}\cite
{Andronov}\cite{Smale}. As emphasized by Flato\cite{Flato} and Faddeev\cite
{Faddeev} the same pattern seems to occur in the fundamental theories of
Nature. In fact, the two most important physical revolutions of this
century, namely the passage from non-relativistic to relativistic and from
classical to quantum mechanics, may be interpreted as the transition from
two unstable theories to two stable ones. In the non-relativistic mechanics
case, one notices that the second cohomology group of the homogeneous
Galileo group does not vanish and the corresponding algebra has a
deformation that leads to the Lorentz algebra which, being semisimple, is
stable. In turn, the transition from classical to quantum mechanics may be
regarded as a deformation of the unstable Poisson algebra of phase-space
functions to the stable Moyal-Vey algebra\cite{Bayen}. I will refer to these
two stabilizing deformations as the $\frac{1}{c}$-deformation and the $\hbar 
$-deformation. The deformed algebras are all equivalent for non-zero values
of $\frac{1}{c}$ and $\hbar $. Hence, relativistic mechanics and quantum
mechanics may be derived from the conditions for stability of their
algebras, but the exact values of the deformation parameters cannot be fixed
by purely algebraic considerations. Instead, the deformation parameters are
fundamental constants to be obtained from experiment. In this sense not only
is deformation theory the theory of stable theories, it is also the theory
that identifies the fundamental constants.

A review of deformation theory and of the transition from non-relativistic
to relativistic and from classical to quantum mechanics as the
deformation-stabilization of two unstable theories is contained in Ref.\cite
{Vilela1}. Also, it is shown there that both deformations may be studied in
the context of finite-dimensional Lie algebras, which is simpler than the
usual treatment of quantum mechanics as a deformation of an
infinite-dimensional algebra of functions. The algebra that results from the 
$\frac{1}{c}$-deformation is the Lorentz algebra and the one coming from the 
$\hbar $-deformation is the Heisenberg algebra. A simple fact in this
construction, which however has non-trivial consequences, is that, to have
both constructions in a finite-dimensional algebra setting, it is essential
to include the coordinates as basic operators in the defining (kinematical)
algebra of relativistic quantum mechanics. The full algebra of relativistic
quantum mechanics will then contain the Lorentz algebra $\{M_{\mu \nu }\}$,
the Heisenberg algebra for the momenta and space-time coordinates $\{p_{\mu
},x_{\nu }\}$ in Minkowski space and also the commutators that define the
vector nature (under the Lorentz group) of $p_{\mu }$ and $x_{\nu }$, namely
\begin{equation}
\begin{array}{rll}
\lbrack M_{\mu \nu },M_{\rho \sigma }] & = & i(M_{\mu \sigma }\eta _{\nu
\rho }+M_{\nu \rho }\eta _{\mu \sigma }-M_{\nu \sigma }\eta _{\mu \rho
}-M_{\mu \rho }\eta _{\nu \sigma }) \\ 
\lbrack M_{\mu \nu },p_{\lambda }] & = & i(p_{\mu }\eta _{\nu \lambda
}-p_{\nu }\eta _{\mu \lambda }) \\ 
\lbrack M_{\mu \nu },x_{\lambda }] & = & i(x_{\mu }\eta _{\nu \lambda
}-x_{\nu }\eta _{\mu \lambda }) \\ 
\lbrack p_{\mu },p_{\nu }] & = & 0 \\ 
\lbrack x_{\mu },x_{\nu }] & = & 0 \\ 
\lbrack p_{\mu },x_{\nu }] & = & i\eta _{\mu \nu }\Im 
\end{array}
\label{1.1}
\end{equation}
with $\eta _{\mu \nu }=(1,-1,-1,-1)$, $c=\hbar =1$ and $\Im $ a unit that
commutes with all the other operators.

One knows that the Lorentz algebra, being semi-simple, is stable and that
each one of the 2-dimensional Heisenberg algebras $\{p_{\mu },x_{\nu }\}$ is
also stable in the non-linear sense discussed in Ref.\cite{Vilela1}. When
the two algebras are combined through the covariance commutators, the
natural question to ask is whether the whole algebra is stable or whether
there are any non-trivial deformations. The answer\cite{Vilela1} is that the
algebra $\Re _{0}=\{M_{\mu \nu },p_{\mu },x_{\mu },\Im \}$ defined by Eqs.(%
\ref{1.1}) is not stable. This is shown by exhibiting a 2-parameter $(\ell
,R)$-deformation of $\Re _{0}$ to a simple algebra $\Re _{\ell ,R}$ which
itself is stable, namely 
\begin{equation}
\begin{array}{rcl}
\lbrack M_{\mu \nu },M_{\rho \sigma }] & = & i(M_{\mu \sigma }\eta _{\nu
\rho }+M_{\nu \rho }\eta _{\mu \sigma }-M_{\nu \sigma }\eta _{\mu \rho
}-M_{\mu \rho }\eta _{\nu \sigma }) \\ 
\lbrack M_{\mu \nu },p_{\lambda }] & = & i(p_{\mu }\eta _{\nu \lambda
}-p_{\nu }\eta _{\mu \lambda }) \\ 
\lbrack M_{\mu \nu },x_{\lambda }] & = & i(x_{\mu }\eta _{\nu \lambda
}-x_{\nu }\eta _{\mu \lambda }) \\ 
\lbrack p_{\mu },p_{\nu }] & = & -i\frac{\epsilon ^{^{\prime }}}{R^{2}}%
M_{\mu \nu } \\ 
\lbrack x_{\mu },x_{\nu }] & = & -i\epsilon \ell ^{2}M_{\mu \nu } \\ 
\lbrack p_{\mu },x_{\nu }] & = & i\eta _{\mu \nu }\Im \\ 
\lbrack p_{\mu },\Im ] & = & -i\frac{\epsilon ^{^{\prime }}}{R^{2}}x_{\mu }
\\ 
\lbrack x_{\mu },\Im ] & = & i\epsilon \ell ^{2}p_{\mu } \\ 
\lbrack M_{\mu \nu },\Im ] & = & 0
\end{array}
\label{1.2}
\end{equation}
$\epsilon $ and $\epsilon ^{^{\prime }}$ being $\pm $ signs. The stable
algebra $\Re _{\ell ,R}$ to which $\Re _{0}$ has been deformed is the
algebra of the 6-dimensional pseudo-orthogonal group with metric $\eta
_{aa}=(1,-1,-1,-1,\epsilon ^{^{\prime }},\epsilon )$ and commutation
relations 
\begin{equation}
\lbrack M_{ab},M_{cd}]=i(-M_{bd}\eta _{ac}-M_{ac}\eta _{bd}+M_{bc}\eta
_{ad}+M_{ad}\eta _{bc})  \label{1.3}
\end{equation}
the correspondence being established by 
\begin{equation}
\begin{array}{lll}
p_{\mu } & = & \frac{1}{R}M_{\mu 4} \\ 
x_{\mu } & = & \ell M_{\mu 5} \\ 
\Im & = & \frac{\ell }{R}M_{45}
\end{array}
\label{1.4}
\end{equation}

To understand the role of the deformation parameters consider first the
Poincar\'{e} subalgebra $P=\{M_{\mu \nu },p_{\mu }\}$ of $\Re _{0}$. It is
well known that already this subalgebra is not stable and may be deformed%
\cite{Flato} \cite{Barut} to the stable simple algebras of the De Sitter
groups O(4,1) or O(3,2). This is the deformation that corresponds to the
parameter $R$. This instability of the Poincar\'{e} algebra is however
physically harmless and well understood. It simply means that flat space is
an isolated point in the set of arbitrarily curved spaces. Faddeev\cite
{Faddeev} points out that Einstein's theory of gravity may be interpreted as
a deformation. This theory is based on curved pseudo Riemannian manifolds.
Therefore, in the set of Riemannian manifolds, Minkowski space is an
isolated point, whereas a generic Riemannian manifold is stable in the sense
that in its neighborhood all spaces are curved. However, as long as the
Poincar\'{e} group is used as the kinematical group of the tangent space to
the space-time manifold, and not as a group of motions in the manifold
itself, it is perfectly consistent to take $R\rightarrow \infty $ and this
deformation goes away.

For the other deformation parameter ($\ell $) there is no reason to imagine
that it should vanish, even in tangent space, if one insists on the
stability paradigm as the guiding principle for theory construction. One is
therefore led to $\Re _{\ell ,\infty }$ as our candidate for a {\it stable
algebra of relativistic quantum mechanics} in the tangent space. The main
features are the non-commutativity of the $x_{\mu }$ coordinates and the
fact that $\Im $, previously a trivial center of the Heisenberg algebra,
becomes now a non-trivial operator. Two constants define this deformation.
One is $\ell $, a fundamental length, the other the sign of $\epsilon $. The
tangent space algebra $\Re _{\ell ,\infty }$ would be the kinematical
algebra appropriate for microphysics. For physics in the large, one might
however use $\Re _{\ell ,R}$ with (finite) $R^{2}$ related to the
gravitational constant $G$.

The idea of modifying the algebra of the space-time components $x_{\mu }$ in
such a way that they become non-commuting operators had already appeared
several times in the physical literature. However, rather than being
motivated (and forced) by stability considerations, the aim of those
proposals was to endow space-time with a discrete structure, to be able, for
example, to construct quantum fields free of ultraviolet divergences.
Sometimes a non-zero commutator was simply postulated, some other times the
motivation was the formulation of field theory in curved spaces. Although
the algebra discussed above is so simple and appears in such a natural way
in the context of deformation theory, it seems that, strangely, it differs
in some way or another from the past proposals. In some schemes, for
example, the coordinates were assumed to be the generators of rotations in a
5-dimensional space with constant negative curvature. This possibility was
proposed long ago by Snyder\cite{Snyder} and the consequences of formulating
field theories in such spaces were extensively studied by Kadishevsky and
collaborators\cite{Kadyshevsky1} \cite{Kadyshevsky2}. The coordinate
commutation relations $[x_{\mu },x_{\nu }]$ are identical to those in (\ref
{1.2}), however, because of the representation chosen for the momentum
operators, the Heisenberg algebra is different and, in particular, $[p_{\mu
},x_{\nu }]$ has non-diagonal terms. Banai\cite{Banai} also proposed a
specific non-zero commutator which only operates between time and space
coordinates, breaking Lorentz invariance. Many other discussions exist
concerning the emergence and the role of discrete or quantum space-time,
which however, in general, do not specify a complete operator algebra\cite
{Das} \cite{Atkinson} \cite{Gudder} \cite{Finkelstein} \cite{Dineykhan} \cite
{Lee} \cite{Prugovecki} \cite{Blokhintsev} \cite{Schild} \cite{Veneziano} 
\cite{t Hooft} \cite{Jackson} \cite{Madore} \cite{Maggiore} \cite{Kempf}.

In $\Re _{\ell ,\infty }$, the fact that $\Im $ becomes a non-trivial
operator changes the structure of the Heisenberg algebra. This has some
consequences on the construction of the state spaces even for
nonrelativistic quantum mechanics. This was partly discussed in Ref.\cite
{Vilela2}. Here the emphasis will be on the study of the geometric structure
of space-time that follows from the $\Re _{\ell ,\infty }$ algebra and on
the nature of quantum fields. In particular the larger set of derivations
that $\Re _{\ell ,\infty }$ possesses has for gauge fields some important
consequences that do not depend on the size of the parameter $\ell $ , but
only on the fact that it is different from zero.

In Sect.2 one collects the basic facts about the non-commutative geometry of
space-time which are implied by the algebra $\Re _{\ell ,\infty }$. In
particular, the role of the elementary sets of the geometry is clarified and
a differential calculus developed. In a general non-commutative geometry
setting \cite{Connes2}, differential calculus cannot be developed through
derivations, because for some algebras there is not enough derivations.
Then, the non-commutative analog of the Dirac operator is used for this
purpose. In the $\Re _{\ell ,\infty }$ case, however, an approach through
derivations is possible. This has the advantage of making the commutative
limit ($\ell \rightarrow 0$) very transparent. Nevertheless, correspondence
with the related Dirac operator approach is also established.

The representation theory of algebras is the basic tool to extract physical
consequences from the non-commutative geometry. This is discussed at length
in Sect.3. In the $\Re _{\ell ,\infty }$ algebra, the usual Heisenberg
algebras are replaced by the algebras of ISO(2) and ISO(1,1). A construction
of a quantum stochastic calculus based on these algebras is sketched in the
Appendix.

Integration in non-commutative space-time is discussed in Sect.4 and finally
Sect.5 is dedicated to the construction of local quantum fields, the main
emphasis being on the non-commutative geometry implications for gauge field
interactions.

\section{The noncommutative space-time geometry}

Every geometrical property of an ordinary (commutative) manifold $M$ may be
expressed as a property of the commutative $C^{*}$-algebra $C_{0}(M)$ of
continuous functions on $M$ vanishing at infinity. For example, there is a
one-to-one correspondence between the characters of $C_{0}(M)$ and the
points of the manifold $M$, regular Borel measures on $M$ correspond to
positive linear functionals on $C_{0}(M)$, complex vector bundles over $M$
are given by the finite projective modules over $C_{0}(M)$, etc. Similarly
in non-commutative geometry one starts from a non-commutative $C^{*}$
-algebra and uses the same correspondence as in the commutative case to
characterize the geometric properties of the non-commutative space\cite
{Connes2}\cite{Connes3}.

In a general representation, the operators in $\Re _{\ell ,\infty }$ are not
bounded operators. However, once a representation of $\Re _{\ell ,\infty }$
is obtained, there are standard ways to construct bounded operators from
unbounded ones, in the universal enveloping algebra of $\Re _{\ell ,\infty }$%
. For example, 
\begin{equation}
\Gamma \rightarrow \exp (i\alpha \Gamma )  \label{2.1}
\end{equation}
or 
\begin{equation}
\Gamma \rightarrow \Gamma (1+\Gamma ^{*}\Gamma )^{-\frac{1}{2}}  \label{2.2}
\end{equation}
and one constructs from the latter, by norm-completion, the associated $C^{*}
$-algebra. Therefore, for simplicity, the discussion of the representations
may be carried out at the $\Re _{\ell ,\infty }$ algebra level, even if the
non-commutative space-time algebra is actually a $C^{*}$-algebra $C_{\Re }$
obtained by the restriction to the bounded operators in the universal
enveloping algebra $U_{\Re }$ of $\Re _{\ell ,\infty }$ and norm completion.

\subsection{The elementary sets of the geometry. Commutative versus
noncommutative space-time}

In the commutative case, the elementary sets of the space-time manifold $M$
(the points with coordinates $x_{\mu }$) are in a one-to-one correspondence
with the character representations of the algebra. In non-commutative case
the elementary sets are the irreducible unitary representations of the
algebra.

In $\Re _{\ell ,\infty }$, the set $A_{M}=\{x_{\mu },M_{\mu \nu }\}$ is the
minimal algebraically closed set that contains the space-time coordinate
operators. It is therefore their representations that define the basic
structure of non-commutative space-time. Therefore it is appropriate to
compare the nature of the representations of the $A_{M}$ algebra for the
commutative and the noncommutative cases. In the commutative case $A_{M}$ is
a Poincar\'{e} algebra and in the noncommutative case it is a DeSitter
algebra (of O(3,2) for $\epsilon =+1$ and of O(4,1) for $\epsilon =-1$). In
the following, for definiteness, I will use $\epsilon =-1$. In the
commutative case the elementary sets $T_{K}$ of the geometry are the
spinless representations of the Poincar\'{e} group corresponding to fixed $%
x_{\mu }x^{\mu }$%
\[
T_{K}=\left\{ x^{\mu
}|(x^{0})^{2}-(x^{1})^{2}-(x^{2})^{2}-(x^{3})^{2}=K\right\} 
\]
From the physical point of view it makes sense to consider these (and not
the points $x_{\mu }$) as the elementary sets of the geometry because each
particular point $x_{\mu }$ in $T_{K}$ is just a particular aspect of the
same event seen in different frames. In the noncommutative case the
elementary sets are the representations of the DeSitter group which reduce
to these Poincar\'{e} group representations in the $\ell \rightarrow 0$
limit.

The correspondence is made very clear by using the explicit representation
of the operators of the $\Re _{\ell ,\infty }$ algebra as differential
operators in a 5-dimensional commutative manifold $M_{5}=\{\xi _{\mu }\}$
with metric $\eta _{aa}=(1,-1,-1,-1,\epsilon )$%
\begin{equation}
\begin{array}{lll}
M_{\mu \nu } & = & i(\xi _{\mu }\frac{\partial }{\partial \xi ^{\nu }}-\xi
_{\nu }\frac{\partial }{\partial \xi ^{\mu }}) \\ 
x_{\mu } & = & \xi _{\mu }+i\ell (\xi _{\mu }\frac{\partial }{\partial \xi
^{4}}-\epsilon \xi ^{4}\frac{\partial }{\partial \xi ^{\mu }})
\end{array}
\label{2.3}
\end{equation}
In $M_{5}$ consider the family of hypersurfaces 
\[
\Gamma _{K}=\{\xi |(\xi ^{0})^{2}-(\xi ^{1})^{2}-(\xi ^{2})^{2}-(\xi
^{3})^{2}-(\xi ^{4})^{2}=K\} 
\]
for $K\in (-\infty ,\infty )$. For each fixed $K$, $\Gamma _{K}$ carries a
representation of the DeSitter group (for $\epsilon =-1$). The intersection
of each $\Gamma _{K}$ with any plane $\xi ^{4}=c$ is a 3-dimensional
hypersurface $T_{K+c^{2}}$ that corresponds to a spinless irreducible
unitary representation of the Poincar\'{e} group. However, because of the $%
(\mu ,4)-$rotations in the $x_{\mu }$ operator (Eq.(\ref{2.3})), it is $%
\Gamma _{K}$ that is irreducible for the $A_{M}$ algebra. Therefore the
elementary sets of the commutative space-time geometry correspond to the $%
T_{K}$ sets and those of noncommutative space-time to the $\Gamma _{K}$ $%
^{^{\prime }}$s. It should however be clear the $T_{K}$ and $\Gamma _{K}$
sets are simply abstract representations of the irreducible representations
of the algebra and their dimensionality should not be confused with the
dimensionality of space-time. The manifold $M_{5}$ is simply the carrier of
the representation of the non-commutative space-time algebra. Space-time is
still defined by the same four $x_{\mu }$ operators operating on the
elementary geometric sets. Fig.1 depicts the structure of the elementary
sets with the $\Gamma _{K}$ $^{^{\prime }}$s, when intersected by the $\xi
^{4}=1$ plane, generating the elementary sets of the pseudo-Euclidean
(commutative) Minkowski geometry (and the intersection with the $\xi ^{0}=0$
plane generating an Euclidean geometry).

\begin{figure}[htb]
\begin{center}
\psfig{figure=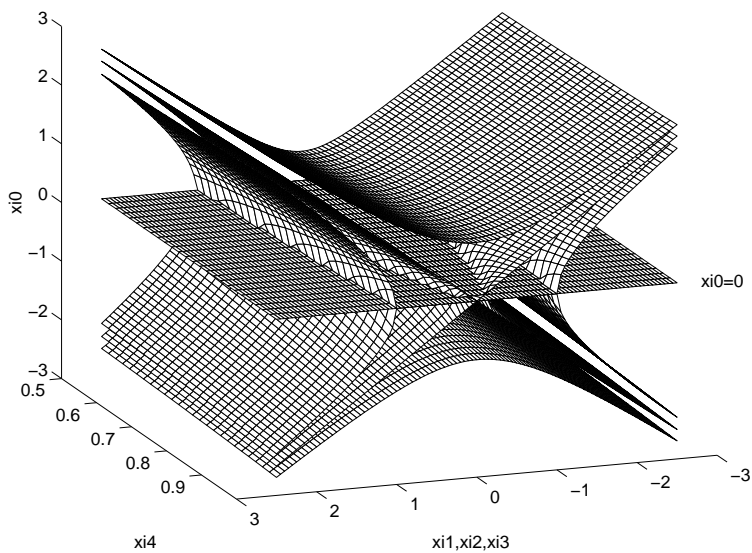,width=10truecm}
\caption[ ]{Representation of elementary sets in space-time geometry:
 commutative versus noncommutative}
\end{center}
\end{figure}

Although not changing the dimensionality of space-time, the $\Gamma _{K}$
sets, as compared to the $T_{K}$'s, have a richer group of motions and, in
particular, a richer set of derivations, as will be seen below.

\subsection{Derivations as vector fields and the differential algebra}

A differential algebra may be defined either by duality from the derivations
of the algebra when a sufficient number of derivations is available or
directly from the triple $\left( H,\pi (C_{\Re }),D\right) $, where $\pi
(C_{\Re })$ is a representation of the $C_{\Re }$ algebra in the Hilbert
space $H$ and $D$ is the Dirac operator. In this latter case the commutator
with the Dirac operator is used to obtain the one-forms. In the most general
non-commutative framework\cite{Connes2} it is not always possible to use the
derivations of the algebra to construct by duality the differential forms.
In fact many algebras have no derivations at all. However when the algebra
has enough derivations it is useful to consider them\cite{Dubois1}\cite
{Dubois2} because the correspondence of the non-commutative geometry notions
to the classical ones becomes very clear. In our case it means to obtain the
usual commutative geometry notions in the limit $\ell \rightarrow 0$. For
this reason the construction through derivations will be used here and the
correspondence to the Dirac commutator approach will be established later on.

Although not essential, the representation of the remaining operators of the 
$\Re _{\ell ,\infty }$ algebra as differential operators on $M_{5}$ provides
an intuitive interpretation of the derivations and is listed below 
\begin{equation}
\begin{array}{lll}
p_{\mu } & = & i\frac{\partial }{\partial \xi ^{\mu }} \\ 
\Im & = & 1+i\ell \frac{\partial }{\partial \xi ^{4}}
\end{array}
\label{2.4}
\end{equation}

The derivations of the algebra play, as in the classical case, the role of
vector fields. The derivations that are considered, to construct by duality
the differential algebra, play only a subsidiary role in identifying the
minimal extension needed when going from the commutative ($\ell =0$) to the
non-commutative case ($\ell \neq 0$). In the end it is the resulting
differential algebra which plays the central role.

The minimal algebraically closed subalgebra that contains the coordinate
operators, $A_{M}=\left\{ x_{\mu },M_{\mu \nu }\right\} $, being semisimple,
it only has inner derivations. In particular because of the commutation
relation $[p_{\mu },x_{\nu }]=i\eta _{\mu \nu }\Im $ and the non-triviality
of the $\Im $ operator, the derivations that correspond to the momentum
operator are not contained in the set of derivations of the enveloping
algebra of $A_{M}$ (Der$\{U_{A_{M}}\}$). Therefore, to obtain enough
derivations, one should consider the full algebra $\Re _{\ell ,\infty }$ and
its generalized enveloping algebra $U_{\Re }$, to which a unit and, for
later convenience, the inverse of $\Im $, are also added. 
\begin{equation}
U_{\Re }=\{x_{\mu },M_{\mu \nu },p_{\mu },\Im ,\Im ^{-1},1\}  \label{2.4a}
\end{equation}
The derivations of $\Re _{\ell ,\infty }$ are the inner derivations plus a
dilation ${\cal D}$ which acts on the generators as follows 
\begin{equation}
\begin{array}{lllll}
\lbrack {\cal D},P_{\mu }] & = & P_{\mu } &  &  \\ 
\lbrack {\cal D},\Im ] & = & \Im  &  &  \\ 
\lbrack {\cal D},M_{\mu \nu }] & = & [{\cal D},x_{\mu }] & = & 0
\end{array}
\label{2.5}
\end{equation}
This may be computed directly or, alternatively, by embedding $\Re _{\ell
,\infty }$ into O(2,5), noticing that this algebra has only inner
derivations and selecting those that operate inside $\Re _{\ell ,\infty }$.
Therefore 
\[
\textnormal{Der}\{\Re _{\ell ,\infty }\}=\{x_{\mu },M_{\mu \nu },p_{\mu },\Im ,%
{\cal D}\}
\]
Any element of the form $a\delta $, where $a\in U_{\Re }$ and $\delta \in $%
Der$\{\Re _{\ell ,\infty }\}$ will be a derivation of the generalized
enveloping algebra $U_{\Re }$. Because of the special role that they play in
the construction of the differential algebra, the derivations corresponding
to $\frac{1}{i}P_{\mu }$ and $\frac{1}{i\ell }\Im $ will be denoted by the
symbols $\partial _{\mu }$ and $\partial _{4}$ to emphasize their role as
elements of Der$\{U_{\Re }\}$ rather than elements of $U_{\Re }$. The action
on the generators is 
\begin{equation}
\begin{array}{lll}
\partial _{\mu }(x_{\nu }) & = & \eta _{\mu \nu }\Im  \\ 
\partial _{4}(x_{\mu }) & = & \ell p_{\mu } \\ 
\partial _{\sigma }(M_{\mu \nu }) & = & \eta _{\sigma \mu }p_{\nu }-\eta
_{\sigma \nu }p_{\mu } \\ 
\partial _{\mu }(p_{\nu }) & = & \partial _{\mu }(\Im )=\partial _{\mu }(1)=0
\\ 
\partial _{4}(M_{\mu \nu }) & = & \partial _{4}(p_{\mu })=\partial _{4}(\Im
)=\partial _{4}(1)=0
\end{array}
\label{2.6}
\end{equation}
In the commutative ($\ell =0$) case a basis for 1-forms is obtained, by
duality, from the set $\{\partial _{\mu }\}$. In the $\ell \neq 0$ case the
set of derivations $\{\partial _{\mu },\partial _{4}\}$ is the minimal set
that contains the usual $\partial _{\mu }$'s, is maximal abelian and is
action closed on the coordinate operators, in the sense that the action of $%
\partial _{\mu }$ on $x_{\nu }$ leads to the operator $\Im $ that
corresponds to $\partial _{4}$ and conversely.

Denote by $V$ the complex vector space of derivations spanned by $\{\partial
_{\mu },\partial _{4}\}$. The algebra of differential forms $\Omega (U_{\Re
})$ is now constructed from the complex $C(V,U_{\Re })$ of multilinear
antisymmetric mappings from $V$ to $U_{\Re }$. For an explicit construction
of $\Omega (U_{\Re })$ use a basis of 1-forms $\{\theta ^{\mu },\theta
^{4}\} $ defined by 
\begin{equation}
\theta ^{a}(\partial _{b})=\delta _{b}^{a},\textnormal{ }a,b\in (0,1,2,3,4)
\label{2.8}
\end{equation}

The operators that are associated to the physical coordinates are just the
four $x_{\mu }$, $\mu \in (0,1,2,3)$. An additional degree of freedom
appears however in the set of derivations. This is not a conjectured extra
dimension but simply a mathematical consequence of the algebraic structure
of $\Re _{\ell ,\infty }$ which, in turn, was a consequence of the
stabilizing deformation of relativistic quantum mechanics. No extra
dimension appears in the set of physical coordinates, because it does not
correspond to any operator in $\Re _{\ell ,\infty }$. However the
derivations in $V$ introduce, by duality, an additional degree of freedom in
the exterior algebra. Therefore all quantum fields that are connections will
pick up some additional components. These additional components, in quantum
fields that are connections, are a consequence of the length parameter $\ell 
$ which does not depend on its magnitude, but only on $\ell $ being $\neq 0$.

A basis for k-forms is $\{\theta ^{a_{1}}\wedge \theta ^{a_{2}}\wedge \cdots
\wedge \theta ^{a_{k}}\}$ where 
\begin{equation}
\theta ^{a_{1}}\wedge \theta ^{a_{2}}\wedge \cdots \wedge \theta ^{a_{k}}=%
\frac{1}{k!}\sum_{P}(-1)^{p}\theta ^{P(a_{1}}\otimes \theta ^{a_{2}}\otimes
\cdots \otimes \theta ^{a_{k})}  \label{2.9}
\end{equation}
$p$ being the parity of the $P$ permutation. A general k-form $\omega \in
\Omega ^{k}(U_{\Re })$ is $\omega =\sum_{a_{1}\cdots a_{k}}b_{a_{1}\cdots
a_{k}}\theta ^{a_{1}}\wedge \theta ^{a_{2}}\wedge \cdots \wedge \theta
^{a_{k}}$ with $b_{a_{1}\cdots a_{k}}\in U_{\Re }$.

Given $\omega _{1}=b_{1}\theta ^{i_{1}}\wedge \cdots \wedge \theta ^{i_{p}}$
and $\omega _{2}=b_{2}\theta ^{j_{1}}\wedge \cdots \wedge \theta ^{j_{k}}$
with $b_{1},b_{2}\in U_{\Re }$, the product is 
\begin{equation}
\omega _{1}\wedge \omega _{2}=b_{1}b_{2}\theta ^{i_{1}}\wedge \cdots \wedge
\theta ^{j_{k}}=(-1)^{pk}\omega _{2}\wedge \omega _{1}+[b_{1},b_{2}]\theta
^{i_{1}}\wedge \cdots \wedge \theta ^{j_{k}}  \label{2.10}
\end{equation}
In the exterior algebra $\Omega (U_{\Re })=\oplus _{p=0}^{\infty }\Omega
^{p}(U_{\Re })$ an exterior derivative is defined as a mapping $d:\Omega
^{p}(U_{\Re })\rightarrow \Omega ^{p+1}(U_{\Re })$ such that 
\begin{equation}
d\omega (\delta _{1},\delta _{2},\cdots ,\delta
_{p+1})=\sum_{k=1}^{p+1}(-1)^{k-1}\delta _{k}\omega (\delta _{1},\cdots ,%
\widehat{\delta _{k}},\cdots ,\delta _{p+1})  \label{2.11}
\end{equation}
where $\delta _{i}\in V$. Notice the absence of commutator terms, in the
definition of the exterior derivative, because the set $V$ is Abelian. $%
d^{2}=0$ follows trivially from the commutation of the derivations.

The elements $\theta ^{a}$ of the 1-form basis do not coincide with $dx_{\mu
}$. Actually 
\begin{equation}
dx_{\mu }=\eta _{\nu \mu }\Im \theta ^{\nu }+\ell p_{\mu }\theta ^{4}
\label{2.12}
\end{equation}
and for the other elements of $\Re _{\ell ,\infty }$%
\begin{equation}
\begin{array}{lll}
dM_{\mu \nu } & = & \left( \eta _{\sigma \mu }p_{\nu }-\eta _{\sigma \nu
}p_{\mu }\right) \theta ^{\sigma } \\ 
dp_{\mu } & = & d\Im =0
\end{array}
\label{2.13}
\end{equation}

We may also define a contraction $i_{\delta }$ as a mapping from $\Omega
^{p}(U_{\Re })$ to $\Omega ^{p-1}(U_{\Re })$ 
\begin{equation}
i_{\delta }\omega (\delta _{1},\cdots ,\delta _{p-1})=\omega (\delta ,\delta
_{1},\cdots ,\delta _{p-1})  \label{2.14}
\end{equation}
with $\omega \in $ $\Omega ^{p}(U_{\Re })$ and $\delta \in V$, and a Lie
derivative $L_{\delta }$%
\begin{equation}
L_{\delta }=di_{\delta }+i_{\delta }d  \label{2.15}
\end{equation}

\subsection{The Dirac operator}

The discussion above was based on the construction of the differential
algebra in non-commutative space-time using the set of derivations $\left\{
\partial _{\mu },\partial _{4}\right\} $. An alternative construction of the
differential algebra in non-commutative geometry follows the method proposed
by Connes\cite{Connes2}, which uses the triple $\left( H,\pi (C_{\Re
}),D\right) $, where $\pi (C_{\Re })$ is a representation of the $C_{\Re }$
algebra in the Hilbert space $H$ and $D$ is the Dirac operator.

Consider the space $L^{2}(M_{5})$ of square-integrable functions on $M_{5}$,
a 5-dimensional pseudo-Riemannian manifold with local metric $\eta
_{ab}=(1,-1,-1,-1,-1)$, and the representation of $C_{\Re }$ on $%
L^{2}(M_{5}) $ induced by Eqs.(\ref{2.3}) and (\ref{2.4}).

The Clifford algebra C(1,4) has, like C(1,3), a representation by 4$\times $%
4 matrices, namely 
\[
\gamma ^{a}=(\gamma ^{0},\gamma ^{1},\gamma ^{2},\gamma ^{3},\gamma
^{4}=i\gamma ^{5}) 
\]
For C(1,4) this is a 2:1 representation because complex C(1,4) is isomorphic
to $M_{16}(C)\times M_{16}(C)$. We may therefore construct over the
pseudo-Riemannian manifold $M_{5}$ a spin bundle with 4-dimensional spinors
with sections defined by 
\begin{equation}
\left( D-m\right) \Psi (x)=0  \label{2.16}
\end{equation}
$D=i\gamma ^{a}\frac{\partial }{\partial x^{a}}$ being the Dirac operator.
The Hilbert space $H$ of the triple $\left( H,\pi (C_{\Re }),D\right) $ is
now the space of square integrable sections of the spin bundle and the
representation $\pi (C_{\Re })$ is the one induced by Eqs.(\ref{2.3}) and (%
\ref{2.4}). The differential algebra may now be constructed by defining
k-forms as the following operators on $H$%
\begin{equation}
\omega =\sum a_{0}\left[ D,a_{1}\right] \cdots \left[ D,a_{k}\right]
\label{2.17}
\end{equation}
with $a_{i}\in C_{\Re }$. Computing the commutators of the Dirac operator
with the elements of $\Re _{\ell ,\infty }$ one obtains 
\begin{equation}
\begin{array}{lll}
\lbrack D,x_{\mu }] & = & i\gamma ^{\nu }\eta _{\nu \mu }\Im +i\gamma
^{4}\ell p_{\mu } \\ 
\lbrack D,M_{\mu \nu }] & = & i\gamma ^{\sigma }\left( \eta _{\sigma \mu
}p_{\nu }-\eta _{\sigma \nu }p_{\mu }\right) \\ 
\lbrack D,p_{\mu }] & = & [D,\Im ]=0
\end{array}
\label{2.18}
\end{equation}
and comparing with (\ref{2.12}-\ref{2.13}) one sees that the same structure
is obtained as with the construction through derivations.

\section{Representations}

Explicit representations of the subalgebras of $\Re _{\ell ,\infty }$ in
spaces of functions are the tools needed to compute the physical
consequences of this type of non-commutative space-time. Here one studies in
detail a few cases, starting from the representations of the 3-dimensional
subalgebra that replaces Heisenberg's algebra.

Consider the subalgebra associated to one-dimensional problems, that is 
\begin{equation}
\begin{array}{lll}
\left[ P,X\right] & = & -i\Im \\ 
\left[ X,\Im \right] & = & i\varepsilon P \\ 
\left[ P,\Im \right] & = & 0
\end{array}
\label{3.1}
\end{equation}
where $P=p\ell $ and $X=\frac{x}{\ell }$ . In these variables, the position
is measured in units of $\ell $ and the momentum in units of $\frac{1}{\ell }
$ .

Let $\varepsilon =-1$. Then (\ref{3.1}) is the algebra of the group of
motions of the plane, ISO(2). Its irreducible representations $T_{r}$ \cite
{Vilenkin} are realized as operators on the space of smooth functions on $%
S^{1}$ with scalar product 
\begin{equation}
\left( f_{1},f_{2}\right) =\frac{1}{2\pi }\int_{0}^{2\pi }f_{1}(\theta
)f_{2}^{*}(\theta )d\theta   \label{3.2}
\end{equation}
the operators being 
\begin{equation}
\begin{array}{lll}
X & = & i\frac{\partial }{\partial \theta } \\ 
P & = & r\sin \theta  \\ 
\Im  & = & r\cos \theta 
\end{array}
\label{3.3}
\end{equation}
The irreducible representations are of two types. For $r\neq 0$ the
irreducible representation $T_{r}$ is infinite dimensional, a convenient
basis being the set of exponentials $\exp \left( -in\theta \right) $%
\begin{equation}
T_{r}=\left\{ e^{-in\theta };n\in Z\right\}   \label{3.4}
\end{equation}
and for $r=0$ the irreducible representations are one-dimensional 
\begin{equation}
T_{0n}=\left\{ e^{-in\theta }\right\}   \label{3.5}
\end{equation}
In $T_{r}$ the operators $V_{+}=iP+\Im $ and $V_{-}=-iP+\Im $ are raising
and lowering operators 
\begin{equation}
\begin{array}{lll}
V_{+}e^{-in\theta } & = & r\textnormal{ }e^{-i(n-1)\theta } \\ 
V_{-}e^{-in\theta } & = & r\textnormal{ }e^{-i(n-1)\theta }
\end{array}
\label{3.6}
\end{equation}
The states $e^{-in\theta }$ being the eigenstates of the position operator $X
$, this one has a discrete spectrum $(=Z$ for $X$ or $=\ell Z$ for $x)$. The
representation with $r=0$ would correspond to a space with a single isolated
point. Therefore it is the representations with $r\neq 0$ that are
physically useful. $\ell $ being the minimal fundamental length, the maximum
momentum, in units of $\frac{1}{\ell }$, is one. Hence, for consistency with
(\ref{3.3}), $r$ might actually be chosen equal to one. The consistency of
this choice will become clear in the study of the harmonic oscillator
spectrum.

For each localized state $e_{n}\thicksim e^{-in\theta }$, $P$ is a random
variable with characteristic function 
\begin{equation}
C(s)=<e_{n},e^{isP}e_{n}>=J_{0}(sr)  \label{3.6a}
\end{equation}
the corresponding probability density being 
\begin{equation}
\begin{array}{lllll}
\mu (P) & = & \frac{1}{\pi }\frac{1}{\sqrt{r^{2}-P^{2}}} &  & \left|
P\right| <r \\ 
& = & 0 &  & \left| P\right| >r
\end{array}
\label{3.6b}
\end{equation}

An elaborate boson calculus, based on the operators of the Heisenberg
algebra, has been developed by several authors \cite{Accardi} \cite{Hudson} 
\cite{Parthasa} \cite{Meyer}. For $\ell \neq 0$ the Heisenberg algebra is
replaced by the algebra of ISO(2). For the calculus based on this algebra it
is useful to represent it as a set of operators acting on a space of
holomorphic functions 
\begin{equation}
\begin{array}{lllll}
X & = & z &  &  \\ 
P & = & \frac{1}{2i}\left( e^{\frac{\partial }{\partial z}}-e^{-\frac{%
\partial }{\partial z}}\right)  & = & \frac{1}{i}\triangle _{-} \\ 
\Im  & = & \frac{1}{2}\left( e^{\frac{\partial }{\partial z}}+e^{-\frac{%
\partial }{\partial z}}\right)  & = & \triangle _{+}
\end{array}
\label{3.6c}
\end{equation}
Let $T_{\alpha }$ be the translation operator by $\alpha $, 
\begin{equation}
T_{\alpha }f(z)=f(z+\alpha )  \label{3.6d}
\end{equation}
Then $\triangle _{-}=\frac{1}{2}\left( T_{1}-T_{-1}\right) $ is a finite
difference operator and $\triangle _{+}=\frac{1}{2}\left(
T_{1}+T_{-1}\right) $ a finite average operator. Therefore, instead of $x$
and $\frac{d}{dx}$ for the Heisenberg algebra, the ISO(2) boson calculus is
based on $z$, $\triangle _{+}$, $\triangle _{-}$ and the relations 
\begin{equation}
\begin{array}{lll}
\left[ \triangle _{-},z\right]  & = & \triangle _{+} \\ 
\left[ \triangle _{+},z\right]  & = & \triangle _{-} \\ 
\left[ \triangle _{+},\triangle _{-}\right]  & = & 0
\end{array}
\label{3.6e}
\end{equation}

On the other hand, (with the choice $\epsilon =-1$) the algebra for the pair 
$P^{0}=\ell p^{0}$, $X^{0}=\frac{x^{0}}{\ell }$ is the algebra of ISO(1,1) 
\begin{equation}
\begin{array}{lll}
\left[ P^{0},X^{0}\right] & = & i\Im \\ 
\left[ X^{0},\Im \right] & = & -iP^{0} \\ 
\left[ P^{0},\Im \right] & = & 0
\end{array}
\label{3.6f}
\end{equation}
In this case the representation, as operators acting on differentiable
functions on the hyperbola, is 
\begin{equation}
\begin{array}{lll}
P^{0} & = & r\sinh \mu \\ 
\Im & = & r\cosh \mu \\ 
X^{0} & = & -i\frac{d}{d\mu }
\end{array}
\label{3.6g}
\end{equation}
Generalized eigenvalues of the time operator are $e^{it\mu }$. Because there
is no $\mu -$periodicity in the hyperbola, there is no discrete quantization
of time, as opposed to the discrete quantization of the space coordinate.
This conclusion, of course, depends on the choice $\epsilon =-1$. The
opposite situation would hold for $\epsilon =+1$.

The main steps of a calculus based on ISO(2) and ISO(1,1) are described in
the Appendix.

\subsection{Modifications to the one-dimensional harmonic oscillator spectrum
}

From the harmonic oscillator Hamiltonian 
\begin{equation}
H=\frac{p^{2}}{2m}+\frac{m\omega ^{2}}{2}x^{2}  \label{3.7}
\end{equation}
using the representation (\ref{3.3}), one obtains the eigenvalue problem 
\begin{equation}
\left( \frac{R^{2}}{2m\ell ^{2}}\sin ^{2}\theta -\frac{m\omega ^{2}\ell ^{2}%
}{2}\frac{d^{2}}{d\theta ^{2}}\right) f(\theta )=Ef(\theta )  \label{3.8}
\end{equation}
Eq.(\ref{3.8}) is a Mathieu equation which one rewrites as 
\begin{equation}
\frac{d^{2}}{d\theta ^{2}}f(\theta )=\left( -a+2q\cos \left( 2\theta \right)
\right) f(\theta )  \label{3.9}
\end{equation}
with 
\begin{equation}
\begin{array}{lll}
a & = & \frac{2E}{2m\omega ^{2}\ell ^{2}}-\frac{R^{2}}{2\ell ^{4}m^{2}\omega
^{2}} \\ 
q & = & -\frac{R^{2}}{4\ell ^{4}m^{2}\omega ^{2}}
\end{array}
\label{3.10}
\end{equation}
and has solutions of four types 
\begin{equation}
\begin{array}{lllllllll}
f_{0} & = & \sum_{m=0}^{\infty }A_{2m+p}\cos (2m+p)\theta &  & p=0 &  & 
\textnormal{or} &  & 1 \\ 
f_{1} & = & \sum_{m=0}^{\infty }B_{2m+p}\sin (2m+p)\theta &  & p=0 &  & 
\textnormal{or} &  & 1
\end{array}
\label{3.11}
\end{equation}
with characteristic values $a$ which are denoted by\cite{Abramowitz} 
\begin{equation}
\begin{array}{llll}
a\circeq a_{r} &  &  & \textnormal{for even periodic solutions} \\ 
a\circeq b_{r} &  &  & \textnormal{for odd periodic solutions}
\end{array}
\label{3.12}
\end{equation}
For small $\ell $ , $q$ is large and one may use the asymptotic form for the
eigenvalues 
\begin{equation}
a_{r}\thicksim b_{r+1}\thicksim -2q+2(2r+1)\sqrt{q}-\frac{\left( 2r+1\right)
^{2}+1}{8}-\frac{\left( 2r+1\right) ^{3}+6r+3}{2^{7}\sqrt{q}}-\cdots
\label{3.13}
\end{equation}
from which, and from (\ref{3.10}), one obtains (using $r=1$) 
\begin{equation}
E_{n}=\left( n+\frac{1}{2}\right) \omega -\frac{\left( 2n+1\right) ^{2}+1}{16%
}m\omega ^{2}\ell ^{2}-\frac{\left( 2n+1\right) ^{3}+3\left( 2n+1\right) }{%
2^{7}}m^{2}\omega ^{3}\ell ^{4}+O(\ell ^{6})  \label{3.14}
\end{equation}
as the corrections to the harmonic oscillator spectrum arising from the $%
\ell \neq 0$ algebra.

\subsection{Barrier problems}

Consider a one dimensional barrier, that is 
\begin{equation}
\begin{array}{llllll}
H & = & P^{2} &  &  & \textnormal{for }x>0 \\ 
H & = & P^{2}+V &  &  & \textnormal{for }x<0
\end{array}
\label{3.15}
\end{equation}
Using the representation (\ref{3.3}), the eigenvalue problem 
\begin{equation}
H\sum_{n\in Z}c_{n}e^{in\theta }=\lambda \sum_{n\in Z}c_{n}e^{in\theta }
\label{3.16}
\end{equation}
leads to the following recurrences 
\begin{equation}
\begin{array}{rlllll}
c_{n-2}+c_{n+2} & = & zc_{n} &  &  & \textnormal{for }n<0 \\ 
c_{n-2}+c_{n+2} & = & z^{^{\prime }}c_{n} &  &  & \textnormal{for }n>0
\end{array}
\label{3.17}
\end{equation}
with 
\begin{equation}
\begin{array}{lll}
z & = & 2-4\lambda \\ 
z^{^{\prime }} & = & 2+4V-4\lambda
\end{array}
\label{3.17a}
\end{equation}
Let $\lambda <V$. For a solution that corresponds to a wave propagating from
the right and being reflected at the barrier, the recurrences in (\ref{3.17}%
) are solved by 
\begin{equation}
\begin{array}{llllll}
c_{n} & = & 0 &  &  & \textnormal{for }n\textnormal{ even} \\ 
c_{n} & = & e^{-\left( n+1\right) \gamma } &  &  & \textnormal{for }n\textnormal
{ odd
and }n\geq 1 \\ 
c_{n} & = & ae^{-in\delta }+be^{in\delta } &  &  & \textnormal{for }n\
textnormal{ odd
and }n\leq -1
\end{array}
\label{3.18}
\end{equation}
With $\gamma =\frac{1}{2}\cosh ^{-1}\left( \frac{z^{^{\prime }}}{2}\right) $
and $\delta =\frac{1}{2}\cos ^{-1}\left( \frac{z}{2}\right) $. the
recurrences are satisfied for $\left| n\right| \geq 3$ and, from the
matching conditions, 
\[
\begin{array}{lll}
c_{-3} & = & zc_{-1}-c_{1} \\ 
c_{-1} & = & z^{^{\prime }}c_{1}-c_{3}
\end{array}
\]
one obtains 
\begin{equation}
\begin{array}{lll}
a & = & \frac{-1}{2i\sin 2\delta }\left\{ \left( z^{^{\prime }}e^{-2\gamma
}-e^{-4\gamma }\right) e^{-i3\delta }-\left( zz^{^{\prime }}e^{-2\gamma
}-ze^{-4\gamma }-e^{-2\gamma }\right) e^{-i\delta }\right\} \\ 
b & = & \frac{1}{2i\sin 2\delta }\left\{ \left( z^{^{\prime }}e^{-2\gamma
}-e^{-4\gamma }\right) e^{i3\delta }-\left( zz^{^{\prime }}e^{-2\gamma
}-ze^{-4\gamma }-e^{-2\gamma }\right) e^{i\delta }\right\}
\end{array}
\label{3.19}
\end{equation}
The constant $\gamma $ controls the decay of the wave function under the
barrier and $\delta $ the wave length of the intensity fluctuations to the
right of the barrier. For small $\lambda $ (that is, small energy), $\delta =%
\frac{1}{2}\sin 2\sqrt{\lambda -\lambda ^{2}}\thickapprox \sqrt{\lambda }$, $%
\sqrt{\lambda }$ being the momentum of the incident wave in units of $\frac{1%
}{\ell }$. Let the momentum in physical units be $p=\frac{\sqrt{\lambda }}{%
\ell }$ . Then, expanding $\delta $%
\begin{equation}
in\delta \thickapprox in\ell p\left\{ 1+\frac{5}{6}\ell ^{2}p^{2}+\cdots
\right\}  \label{3.20}
\end{equation}
The conclusion is that the intensity fluctuations to the right of the
barrier have a wave length smaller than the inverse momentum, the leading
correction factor being $\frac{5}{6}\ell ^{2}p^{2}$.

\subsection{Diffraction}

The representation (\ref{3.3}) also provides all the required framework to
compute the effects of the fundamental length $\ell $ on the diffraction
experiments. Let a matter wave pass through a slit of width $L=2N\ell $. The
wave function at the slit may be represented by 
\begin{equation}
\Psi _{L}=\frac{1}{\sqrt{2N}}\sum_{n=-N}^{N}e^{in\theta }  \label{3.21}
\end{equation}
(Generalized) eigenstates of the momentum ($P=\sin \theta $) in the $%
e^{in\theta }$ basis are 
\begin{equation}
\phi _{k}=\frac{\left( 1-k^{2}\right) ^{-\frac{1}{4}}}{\sqrt{2\pi }}%
\sum_{n}e^{-in\sin ^{-1}(k)}e^{in\theta }  \label{3.22}
\end{equation}
the factor $\left( 1-k^{2}\right) ^{-\frac{1}{4}}$ being included to insure
a normalization $\delta \left( k-k^{^{\prime }}\right) $ of the momentum
states.

Computing the projection 
\begin{equation}
\begin{array}{lll}
\sqrt{4\pi N}\left\langle \phi _{k}\mid \Psi _{L}\right\rangle  & = & \left(
1-k^{2}\right) ^{-\frac{1}{4}}\frac{1}{N}\sum_{n=-N}^{N}e^{in\sin ^{-1}(k)}
\\ 
& = & \left( 1-k^{2}\right) ^{-\frac{1}{4}}\frac{1}{2N}\left\{
1+2\sum_{n=1}^{N}T_{n}(\sqrt{1-k^{2}})\right\} 
\end{array}
\label{3.23}
\end{equation}
where $T_{n}$ is a Chebyshev polynomial.

With the transverse momentum $q$ in physical units, $q=\frac{k}{\ell }$,
and, for large $N$, approximating the sum in (\ref{3.23}) by an integral one
obtains 
\begin{equation}
\begin{array}{lll}
4\pi N\left| \left\langle \phi _{k}\mid \Psi _{L}\right\rangle \right| ^{2}
& \simeq & \left( 1-k^{2}\right) ^{-\frac{1}{2}}\left( \frac{\sin \left(
N\sin ^{-1}(k)\right) }{N\sin ^{-1}(k)}\right) ^{2} \\ 
& \simeq & \left( 1-\ell ^{2}q^{2}\right) ^{-\frac{1}{2}}\left( \frac{2}{Lq}%
\sin \left( \frac{Lq}{2}(1+\frac{1}{6}\ell ^{2}q^{2}+\cdots )\right) \right)
^{2}
\end{array}
\label{3.24}
\end{equation}
meaning that for large transverse momentum (large diffraction angles) the
separation between the diffraction rings becomes smaller.

The same result could have been obtained by studying the distribution of the
random variable with characteristic function 
\begin{equation}
\left\langle \psi _{L},e^{isP}\psi _{L}\right\rangle =\frac{1}{2N}%
\sum_{k=-2N}^{2N}\left( 2N-\left| k\right| +1\right) J_{k}(s)  \label{3.24a}
\end{equation}

\subsection{Stochastic processes in noncommutative space-time}

In the Appendix, a quantum stochastic calculus is developed, based on
ISO(1,1) and ISO(2), that are the algebraic structures of the operator sets $%
\left\{ X^{0},P^{0},\Im \right\} $ and $\left\{ X^{i},P^{i},\Im \right\} $.
The stochastic processes constructed there, are sums of independent
identically distributed random variables. Therefore, the ''time'' of the
process is simply the continuous parameter that labels the probability
convolution semigroup. If however time is an operator that satisfies well
defined algebraic relations with the other observables, as in the (\ref{1.2}%
), the construction has to be done in a different manner. The notion of
filtration, in particular, cannot be obtained simply by a splitting of the
indexing space $h$. It must be replaced by a construction of the spaces of
eigenstates of the time operator. Physically the treatment of time as a
parameter still makes sense if the time scale of the processes is slow
(Remember that $x_{0}=ct$ and then $[t,x_{i}]=\frac{i\ell ^{2}}{c}M_{oi}$).
However for processes with a fast time scale, a construction where time is
treated as an operator is needed.

To describe time-dependent processes one needs at least one space and one
time coordinate. Therefore, the minimal algebra is $\left\{
x^{0},x^{1},M^{01},p^{0},p^{1},\Im \right\} $ which, for $\varepsilon =-1$,
is the algebra of $ISO(2,1)$, the group of motions of pseudo-Euclidean
3-space $E_{2,1}$. Representations may be realized on the space of functions
on the double- or single-sheeted hyperboloid $H_{+}^{2}$ or $H_{-}^{2}$ and
on the cone $C^{2}$ , with coordinates, respectively 
\begin{equation}
\left\{ 
\begin{array}{lll}
\xi _{1} & = & \sinh \mu \sin \theta  \\ 
\xi _{2} & = & \sinh \mu \cos \theta  \\ 
\xi _{3} & = & \cosh \mu 
\end{array}
\right\}   \label{3.25}
\end{equation}
\begin{equation}
\left\{ 
\begin{array}{lll}
\xi _{1} & = & \cosh \mu \sin \theta  \\ 
\xi _{2} & = & \cosh \mu \cos \theta  \\ 
\xi _{3} & = & \sinh \mu 
\end{array}
\right\}   \label{3.26}
\end{equation}
\begin{equation}
\left\{ 
\begin{array}{lll}
\xi _{1} & = & r\sin \theta  \\ 
\xi _{2} & = & r\cos \theta  \\ 
\xi _{3} & = & r
\end{array}
\right\}   \label{3.27}
\end{equation}
Here representations on the space of functions on the upper sheet of the
cone $C^{2}$ will be chosen. The reason for this choice is to have positive
energy but no minimal nonzero energy. Then 
\begin{equation}
\begin{array}{lll}
\ell p^{1} & = & r\sin \theta  \\ 
\Im  & = & r\cos \theta  \\ 
\ell p^{0} & = & r \\ 
\frac{x^{0}}{\ell } & = & -i\left\{ -\sin \theta \frac{\partial }{\partial
\theta }+r\cos \theta \frac{\partial }{\partial r}\right\}  \\ 
\frac{x^{1}}{\ell } & = & i\frac{\partial }{\partial \theta } \\ 
M^{01} & = & -i\left\{ \cos \theta \frac{\partial }{\partial \theta }+r\sin
\theta \frac{\partial }{\partial r}\right\} 
\end{array}
\label{3.28}
\end{equation}
acting on functions $f\left( r,\theta \right) $ on $C^{2}$,
square-integrable for the measure $drd\theta $. Hermitean symmetry of the
operators is obtained if either $r^{\frac{1}{2}}f(r,\theta )\rightarrow 0$
when $r\rightarrow \infty $ or $f(r,\theta )=r^{-\frac{1}{2}+i\rho }g(\theta
)$. The last case corresponds to the principal series of $SO(2,1)$
representations.

One sees that the time and the space coordinates are noncommuting operators.
Therefore, when describing a process, time cannot be simply considered a
c-number parameter. Instead, to describe, for example, a stochastic process
that at each fixed time may be sampled to find out the value of the space
variable, what one has to do is to find the subspaces of time eigenvectors,
corresponding to each fixed eigenvalue $t$. Then, in each such subspace, one
has to find the possible values of $x$ and their probabilities. If no
further constraints are imposed on the values of $x$, this will be the
analog of Brownian motion in the noncommutative one-time one-space setting.

The eigenvectors of the time operator in (\ref{3.28}) are obtained from 
\begin{equation}
i\left\{ \sin \theta \frac{\partial }{\partial \theta }-r\cos \frac{\partial 
}{\partial r}\right\} f_{t}(r,\theta )=\frac{t}{\ell }f_{t}(r,\theta )
\label{3.29}
\end{equation}
the solution being 
\begin{equation}
f_{t}(r,\theta )=\left( \cot \frac{\theta }{2}\right) ^{i\frac{t}{\ell }%
}g(r\sin \theta )  \label{3.30}
\end{equation}
with $g(r\sin \theta )$ an arbitrary function of $r\sin \theta $. Now one
considers the spectrum of possible values of the space coordinate $x$ in
each one of the subspaces spanned by the functions $f_{t}(r,\theta ).$ The
projection on the $e^{-in\theta }$ eigenstate of the position operator $x$
(corresponding to the position $n\ell $) is 
\begin{equation}
c_{-n}(t)=\frac{1}{2\pi }\int d\theta e^{in\theta }\left( \cot \frac{\theta 
}{2}\right) ^{i\frac{t}{\ell }}g(r\sin \theta )  \label{3.31}
\end{equation}
For a process that starts from $x=0$ at time $t=0$ it should be $c_{0}(0)=1$
and $c_{n}(0)=0$ for $n\neq 0$. Therefore for such a process $g(r\sin \theta
)=$constant. Strictly speaking a constant function is outside the $L^{2}$
domain of the operator and therefore one should consider the operator as
acting in the generalized functions space of a Gelfand triplet. With the
choice $f_{t}(r,\theta )=\left( \cot \frac{\theta }{2}\right) ^{i\frac{t}{%
\ell }}$ one obtains, by computing the integral (\ref{3.31}) 
\begin{equation}
c_{0}(t)=e^{-\frac{\pi }{2}\frac{t}{\ell }}  \label{3.32}
\end{equation}
For the other coefficients, they are more conveniently obtained by solving (%
\ref{3.29}) in a $e^{-in\theta }$ basis, 
\[
f_{t}(r,\theta )=r^{\sigma }\sum_{n\in Z}c_{n}e^{in\theta }
\]
which leads to the recurrence 
\begin{equation}
(n-1-\sigma )c_{n-1}-(n+1+\sigma )c_{n+1}+i\frac{2t}{\ell }c_{n}=0
\label{3.33}
\end{equation}
For $\sigma =0$, using a recurrence relation for hypergeometric functions
one obtains $c_{n}\thicksim \frac{1}{n}F(-n,-i\frac{t}{\ell };0;2)$. Then 
\begin{equation}
\begin{array}{llll}
c_{n}(t)=\frac{1}{2}e^{in\frac{\pi }{2}}e^{-\frac{\pi }{2}\frac{t}{\ell }%
}P_{n}^{0}(\frac{t}{\ell };\frac{\pi }{2}) &  & \textnormal{for } & n\neq 0
\end{array}
\label{3.34}
\end{equation}
$P_{n}^{0}(\frac{t}{\ell };\frac{\pi }{2})$ being Pollaczek-Meixner
polynomials with generating function 
\begin{equation}
\left( 1-it\right) ^{ix}\left( 1+it\right) ^{-ix}=\sum_{n=0}^{\infty
}P_{n}^{0}(x;\frac{\pi }{2})t^{n}  \label{3.35}
\end{equation}

Hence, without further restrictions on the dynamics, the process that at $t=0
$ starts from $x=0$, has a probability to be found at $\pm n\ell $ at time $t
$ equal to 
\begin{equation}
P(\pm n\ell ,0;\frac{t}{\ell })=\frac{1}{4}e^{-\pi \frac{t}{\ell }}\left(
P_{n}^{0}(\frac{t}{\ell };\frac{\pi }{2})\right) ^{2}  \label{3.36}
\end{equation}

\subsection{Higher dimensional representations}

The full algebra $\Re _{\ell ,\infty }$ , described in Sect.2, is isomorphic
to the algebra of $ISO(4,1)$, the group of motions of the pseudo-Euclidean $%
5 $-space $E_{4,1}$. Consequently, as pointed out in Sect.2, a
representation may be obtained in the form of differential operators in a
5-dimensional commutative manifold. Alternatively, as for the lower
dimension subalgebras treated before, a representation is obtained in the
space of functions on the upper sheet of the cone $C^{4}$, with coordinates 
\begin{equation}
\begin{array}{lll}
\xi _{1} & = & r\sin \theta _{3}\sin \theta _{2}\sin \theta _{1} \\ 
\xi _{2} & = & r\sin \theta _{3}\sin \theta _{2}\cos \theta _{1} \\ 
\xi _{3} & = & r\sin \theta _{3}\cos \theta _{2} \\ 
\xi _{4} & = & r\cos \theta _{3} \\ 
\xi _{5} & = & r
\end{array}
\label{3.37}
\end{equation}
the invariant measure for which the functions are square-integrable being 
\begin{equation}
d\nu (r,\theta _{i})=r^{2}\sin ^{2}\theta _{3}\sin \theta _{2}drd\theta
_{1}d\theta _{2}d\theta _{3}  \label{3.37a}
\end{equation}

On these functions the operators of $\Re _{\ell ,\infty }$ act as follows 
\begin{equation}
\begin{array}{lll}
\ell p^{0} & = & r \\ 
\Im & = & r\cos \theta _{3} \\ 
\ell p^{1} & = & r\sin \theta _{3}\cos \theta _{2} \\ 
\ell p^{2} & = & r\sin \theta _{3}\sin \theta _{2}\cos \theta _{1} \\ 
\ell p^{3} & = & r\sin \theta _{3}\sin \theta _{2}\sin \theta _{1} \\ 
M^{23} & = & -i\frac{\partial }{\partial \theta _{1}} \\ 
M^{12} & = & -i\left( \cos \theta _{1}\frac{\partial }{\partial \theta _{2}}%
-\sin \theta _{1}\cot \theta _{2}\frac{\partial }{\partial \theta _{1}}%
\right) \\ 
M^{31} & = & i\left( \sin \theta _{1}\frac{\partial }{\partial \theta _{2}}%
+\cos \theta _{1}\cot \theta _{2}\frac{\partial }{\partial \theta _{1}}%
\right) \\ 
\frac{x^{0}}{\ell } & = & -i\left( -\sin \theta _{3}\frac{\partial }{%
\partial \theta _{3}}+r\cos \theta _{3}\frac{\partial }{\partial r}\right)
\\ 
\frac{x^{1}}{\ell } & = & i\left( \cos \theta _{2}\frac{\partial }{\partial
\theta _{3}}-\sin \theta _{2}\cot \theta _{3}\frac{\partial }{\partial
\theta _{2}}\right) \\ 
\frac{x^{2}}{\ell } & = & i\left( \cos \theta _{1}\sin \theta _{2}\frac{%
\partial }{\partial \theta _{3}}+\cos \theta _{1}\cos \theta _{2}\cot \theta
_{3}\frac{\partial }{\partial \theta _{2}}-\frac{\sin \theta _{1}}{\sin
\theta _{2}}\cot \theta _{3}\frac{\partial }{\partial \theta _{1}}\right) \\ 
\frac{x^{3}}{\ell } & = & i\left( \sin \theta _{1}\sin \theta _{2}\frac{%
\partial }{\partial \theta _{3}}+\sin \theta _{1}\cos \theta _{2}\cot \theta
_{3}\frac{\partial }{\partial \theta _{2}}+\frac{\cos \theta _{1}}{\sin
\theta _{2}}\cot \theta _{3}\frac{\partial }{\partial \theta _{1}}\right) \\ 
M^{01} & = & i\left( \frac{\sin \theta _{2}}{\sin \theta _{3}}\frac{\partial 
}{\partial \theta _{2}}-\cos \theta _{2}\cos \theta _{3}\frac{\partial }{%
\partial \theta _{3}}-r\cos \theta _{2}\sin \theta _{3}\frac{\partial }{%
\partial r}\right) \\ 
M^{02} & = & -i\left( 
\begin{array}{c}
\frac{\cos \theta _{1}\cos \theta _{2}}{\sin \theta _{3}}\frac{\partial }{%
\partial \theta _{2}}-\frac{\sin \theta _{1}}{\sin \theta _{2}\sin \theta
_{3}}\frac{\partial }{\partial \theta _{1}}+\cos \theta _{1}\sin \theta
_{2}\cos \theta _{3}\frac{\partial }{\partial \theta _{3}}+ \\ 
r\cos \theta _{1}\sin \theta _{2}\sin \theta _{3}\frac{\partial }{\partial r}
\end{array}
\right) \\ 
M^{03} & = & -i\left( 
\begin{array}{c}
\frac{\sin \theta _{1}\cos \theta _{2}}{\sin \theta _{3}}\frac{\partial }{%
\partial \theta _{2}}+\frac{\cos \theta _{1}}{\sin \theta _{2}\sin \theta
_{3}}\frac{\partial }{\partial \theta _{1}}+\sin \theta _{1}\sin \theta
_{2}\cos \theta _{3}\frac{\partial }{\partial \theta _{3}}+ \\ 
r\sin \theta _{1}\sin \theta _{2}\sin \theta _{3}\frac{\partial }{\partial r}
\end{array}
\right)
\end{array}
\label{3.38}
\end{equation}

This is the appropriate representation to generalize to higher dimensions
the construction of processes carried out in the previous subsections.

\section{Integration in noncommutative space-time}

Here one has to distinguish two cases. Because the energy-momentum operators 
$\left\{ p^{0},p^{i}\right\} $ are a commuting set, integration in momentum
space is the usual commutative Lebesgue integration. However the domain of
integration must be consistent with the structure of the algebra of
observables. From the representation (\ref{3.38}) one sees that, when $%
r,\theta _{1},\theta _{2},\theta _{3}$ are diagonalized, integration over
momentum space corresponds to 
\begin{equation}
\int F\left( p^{\mu }\right) d\nu (r,\theta _{i})  \label{4.1}
\end{equation}
$d\nu (r,\theta _{i})$ being the invariant measure defined in (\ref{3.37a})
and the $p^{\mu }$ in the functional $F$ being replaced by their
representations in (\ref{3.38}).

Integration over configuration space, however, differs from Lebesgue
integration because $\{x^{\mu }\}$ is not a commuting set.

In the commutative case an integral 
\begin{equation}
\int f(x)d\nu (x)  \label{4.2}
\end{equation}
has the following algebraic interpretation: In a representation where $x$ is
diagonalized $f(x)$ is the diagonal element $(x,fx)$ and the integral (\ref
{4.2}) is a weighed trace, with the weights assigned to each eigenvalue by
the measure $\nu (x)$. For compact operators a non-commutative integration
theory has been developed with the integral replaced by the Dixmier trace.
Infinitesimals of order $1$ are compact operators with eigenvalues $\mu
=O(n^{-1})$ as $n\rightarrow \infty $. Then, given the sequence 
\begin{equation}
\gamma _{N}=\frac{1}{\log N}\sum_{0}^{N-1}\mu _{n}  \label{4.3}
\end{equation}
there is a linear form $\lim_{\omega }$ on the space $\ell ^{\infty }$ of
bounded sequences which satisfies the properties of linearity and scale
invariance needed to interpret it as an algebraic substitute for the notion
of integral. This form is the Dixmier trace which, if the sequence (\ref{4.3}%
) converges, coincides with its limit.

The coordinate operators are not compact operators. Therefore, when
constructing the non-commutative version of integration over configuration
space, the question of the regularization factor in the trace should be
carefully analysed. Consider first integration in one space variable. As
discussed in Sect.3, the spectrum of $x$ is $\{n\ell :n\in Z\}$ . Therefore
the trace is a sum over $n$%
\[
\lim_{N\rightarrow \infty }\sum_{n=-N}^{N}F\left( x\right)
=\lim_{N\rightarrow \infty }\sum_{n=-N}^{N}F\left( n\ell \right) 
\]
The question is whether one needs a $N-$dependent regularization factor, to
interpret this trace as the integral (like the $\log N$ in the Dixmier
trace). Comparing 
\[
\int^{N}\frac{1}{x}dx\sim \log N
\]
with 
\[
\sum_{1}^{N}\frac{1}{n}=C+\log N+O\left( \frac{1}{N}\right) 
\]
the conclusion is that, in this case, the trace itself has the same
singularity structure as the corresponding continuous integral.. Therefore
it seems consistent to simply use the trace, without any regularization
factor, as the definition of the integral. This is carried over to
configuration space integration on the 4-dimensional case by defining an
orthonormal basis 
\[
\left| \overrightarrow{n}m\right) =\left( \frac{1}{2\pi }\right) ^{\frac{3}{2%
}}e^{i\overrightarrow{n}\cdot \overrightarrow{\theta }}H_{m}(r)
\]
for functions on the cone $C^{4}$,where $\overrightarrow{n}=\left(
n_{1},n_{2},n_{3}\right) $, $\overrightarrow{\theta }=\left( \theta
_{1},\theta _{2},\theta _{3}\right) $ and $H_{m}(r)$ is an Hermite
polynomial. Integration is then defined by the trace. 
\begin{equation}
\sum_{n_{i},m}\left( \overrightarrow{n}m,F\left( x^{\mu }\right) 
\overrightarrow{n}m\right)   \label{4.4}
\end{equation}
being understood that the $x^{\mu }$ in (\ref{4.4}) are represented by the
operators in (\ref{3.38}).

\section{Quantum fields in noncommutative space-time}

\subsection{Local free fields}

In the $\Re _{\ell ,\infty }$ algebra, $\left\{ P^{\mu },\Im \right\} $ is a
commuting set. Therefore a complete set of eigenstates of the momentum may
be constructed and, in momentum-space, calculations may be carried out as in
the commutative case. However quantum fields over space-time are also
needed, to construct local interactions. Because the momenta and the
coordinates are not Heisenberg dual, the usual Fourier transform cannot be
used to construct local fields. This is then replaced by the following
construction:

Given a representation where the operators $\left\{ P^{\mu },\Im \right\} $
act as multiplicative operators in a space of functions (Sect.3), $\Im ^{-1}$
is also a well defined multiplicative operator. Then, a set that obeys
Heisenberg commutation relations with $\left\{ P^{\mu }\right\} $ is the set 
$\left\{ y^{\mu }=\frac{1}{2}\{x^{\mu },\Im ^{-1}\}_{+}\right\} $, where $%
\{...\}_{+}$ denotes the anticommutator. This set may be used to construct
local fields, from the momentum space states, by Fourier transform. Notice
however that $\left\{ y^{\mu }\right\} $ is still a non-commuting set and
the non-commutative nature of the geometry is fully preserved.

In the commutative case, fields are sections of vector bundles $E$ over the
configuration space $M$ and the space of sections is a representation space
for the algebra of functions on the base manifold (more precisely a
projective module). Moreover it is known that for compact $M$ there is a
one-to-one correspondence between vector bundles and finite projective
modules over the space $C(M)$ of continuous functions on $M$ \cite{Serre}%
\cite{Swan}. This is the correspondence that provides a generalization to
the non-commutative case. The notion is carried over to the non-commutative
case as follows. Let 
\begin{equation}
E_{\pi }=\{\psi \in U_{\Re }\otimes U_{\Re }\otimes \cdots \otimes U_{\Re
}:\pi \psi =\psi \}  \label{5.1}
\end{equation}
The non-commutative version of a section (n-component quantum field) is an
element of the n-fold tensor product of the generalized enveloping algebra $%
U_{\Re }$ defined in (\ref{2.4a}) (Sect.2.2), restricted by the projector
relation $\pi \psi =0$. $(\pi -1)\psi =0$ is the equivalent of a field
equation.

To fully appreciate the similarities and differences to the commutative case
one follows a construction as close as possible to the commutative one. For
this purpose one profits from the commutative nature of the energy-momentum
operator set. In $U_{\Re }$ one has the relation 
\begin{equation}
\left[ p_{\mu },e^{\frac{i}{2}k_{\nu }\left\{ x^{\nu },\Im ^{-1}\right\}
_{+}}\right] =ik_{\mu }e^{\frac{i}{2}k_{\nu }\left\{ x^{\nu },\Im
^{-1}\right\} _{+}}  \label{5.2}
\end{equation}
Therefore $\phi \in U_{\Re }$ given by 
\begin{equation}
\phi =\int d^{4}k\delta (k^{2}-m^{2})\left\{ a_{k}e^{\frac{i}{2}k_{\nu
}\left\{ x^{\nu },\Im ^{-1}\right\} _{+}}+b_{k}^{*}e^{-\frac{i}{2}k_{\nu
}\left\{ x^{\nu },\Im ^{-1}\right\} _{+}}\right\}   \label{5.3}
\end{equation}
satisfies the (projection) equation

\begin{equation}
\lbrack P_{\mu },[P^{\mu },\phi ]]-m^{2}\phi =0  \label{5.4}
\end{equation}
and is a free scalar field in noncommutative space-time. The local field $%
\phi $ is an element of the enveloping algebra $U_{\Re }$. Therefore powers
of $\phi $, multiplication and the action of the derivations being well
defined, the non-commutative version of local interactions is also well
defined.

Similarly free spinor fields may be defined by 
\begin{equation}
\psi =\int d^{4}k\delta (k^{2}-m^{2})\left\{ b_{k}u_{k}e^{-\frac{i}{2}k_{\nu
}\left\{ x^{\nu },\Im ^{-1}\right\} _{+}}+d_{k}^{*}v_{k}e^{\frac{i}{2}k_{\nu
}\left\{ x^{\nu },\Im ^{-1}\right\} _{+}}\right\}   \label{5.5}
\end{equation}

\begin{equation}
\psi \in U_{\Re }:D\psi -m\psi =0  \label{7.10}
\end{equation}
$D$ being the Dirac operator defined before (Sect.2)

\subsection{Gauge fields}

Consider now gauge fields in the non-commutative space-time context. Gauge
fields in the commutative case are Lie algebra-valued connections.

In the simplest case consider a right $U_{\Re }$-module generated by $1$. 
\begin{equation}
E=\{1a;\bigskip \ a\in U_{\Re }\}  \label{8.23}
\end{equation}
A connection is a mapping $\nabla :E\rightarrow E\otimes \Omega ^{1}(U_{\Re
})$ such that 
\begin{equation}
\nabla (\chi a)=\chi da+\nabla (\chi )a  \label{8.24}
\end{equation}
$\chi \in E$, $a\in U_{\Re }$. For each derivation $\delta _{i}\in V$ the
connection defines a mapping $\nabla _{\delta _{i}}:E\rightarrow E$. Because
of Eq.(\ref{8.24}), if one knows how the connection acts on the algebra unit 
$1$, one has the complete action. Define 
\begin{equation}
\nabla (1)\doteq A=A_{i}\theta ^{i},\bigskip \ A_{i}\in U_{\Re }
\label{8.25}
\end{equation}

A gauge transformation is a unitary element ($U^{*}U=1$) acting on $E$. Such
unitary elements exist in the $C^{*}$-algebra formed from the elements of
the enveloping algebra by the standard techniques.

Let $\phi \in E$ be a scalar field. Then 
\begin{equation}  \label{8.26}
\nabla (\phi )=d\phi +\nabla (1)\phi
\end{equation}
Acting on $\nabla (\phi )$ with a unitary element 
\begin{equation}  \label{8.27}
U\nabla (\phi )=Ud(U^{-1}U\phi )+U\nabla (1)U^{-1}U\phi =d(U\phi
)+\{U(dU^{-1})+U\nabla (1)U^{-1}\}U\phi =\nabla ^{^{\prime }}(U\phi )
\end{equation}
Therefore the gauge field transforms as follows under a gauge transformation 
\begin{equation}  \label{8.28}
\nabla (1)\rightarrow U(dU^{-1})+U\nabla (1)U^{-1}
\end{equation}
Notice that the second term does not vanish because of the non-commutativity
of $U_\Re $.

The connection is extended to a mapping $E\otimes \Omega (U_\Re )\rightarrow
E\otimes \Omega (U_\Re )$ by 
\begin{equation}  \label{8.29}
\nabla (\phi \alpha )=\nabla (\phi )\alpha +\phi d\alpha
\end{equation}
$\phi \in E$ and $\alpha \in \Omega (U_\Re )$. We may now compute $\nabla
^2(1)$%
\begin{equation}  \label{8.30}
\begin{array}{c}
\nabla ^2(1)=\nabla (1A_i\theta ^i)=\nabla (1A_i)\theta ^i+1A_id\theta ^i \\ 
=1dA_i\theta ^i+\nabla (1)A_i\theta ^i+1A_id\theta ^i \\ 
=\partial _j(A_i)\theta ^j\wedge \theta ^i+A_jA_i\theta ^j\wedge \theta ^i
\end{array}
\end{equation}
Therefore given an electromagnetic potential $A=A_i\theta ^i$ ($A_i\in U_\Re 
$) the corresponding electromagnetic field is $F_{ij}\theta ^i\wedge \theta
^j$ where 
\begin{equation}  \label{8.31}
F_{ij}=\partial _i(A_j)-\partial _j(A_i)+[A_i,A_j]
\end{equation}
$F_{ij}\in U_\Re $.

Unlike the situation in commutative space-time, the commutator term does not
vanish and pure electromagnetism is no longer a free theory, because of the
quadratic terms in $F_{ij}$. Notice also that the indices in the connections
(\ref{8.25}) and gauge fields (\ref{8.31}) run over (0,1,2,3,4).

To construct an action for the gauge fields an integration on forms is
needed. Because of the structure of the derivation algebra $\Omega
^{5}(U_{\Re })$ is generated by $\theta ^{0}\wedge \theta ^{1}\wedge \theta
^{2}\wedge \theta ^{3}\wedge \theta ^{4}$. Therefore, given an arbitrary
element of $\Omega ^{5}(U_{\Re })$%
\begin{equation}
A=a\smallskip \ \theta ^{0}\wedge \theta ^{1}\wedge \theta ^{2}\wedge \theta
^{3}\wedge \theta ^{4}  \label{8.32}
\end{equation}
we define 
\begin{equation}
\int A=\textnormal{Tr}(a)  \label{8.33}
\end{equation}
By Tr we mean the trace in the sense discussed in Sect.4, if a basis for the
representation of $U_{\Re }$, as operators acting on a space of functions on
the cone $C^{4}$, is used. As discussed before, if this representation is
used, the trace has the same singularity structure as the corresponding
commutative integral. For other representations however, a regularizing
factor (as in the Dixmier trace) may have to be used.

To construct an action for the electromagnetic field consider a diagonal
metric $\eta _{ab}=(1,-1,-1,-1,-1)$ and construct 
\begin{equation}
G=G_{knl}\theta ^{k}\wedge \theta ^{n}\wedge \theta ^{l}  \label{8.34}
\end{equation}
where $G_{knl}=\epsilon _{\cdot \cdot knl}^{ij}F_{ij}\in U_{\Re }$. The
action $S_{A}$ is obtained from the trace of $F\wedge G$%
\begin{equation}
S_{A}=\textnormal{Tr}\{F_{ab}F^{ab}\}=\textnormal{Tr}\{F_{\mu \nu }F^{\mu \nu }+2F_{4\mu
}F^{4\mu }\}  \label{8.35}
\end{equation}
$\mu ,\nu \in (0,1,2,3)$.

In conclusion one finds:

- Additional fields ($A_4,F_{4\mu }$) in non-commutative electromagnetism,

- Non-linear terms in $F_{ab}$,

- Additional terms in the action.

The existence of additional field components is not associated to extra
dimensions or a multi-sheeted nature for space-time. They appear only
because of the existence of the derivation $\partial _{4}$. This however
operates inside the algebra of the usual physical observables, namely 
\begin{equation}
\partial _{4}(x_{\mu })=\ell P_{\mu }  \label{8.36}
\end{equation}

The gauge fields in non-commutative space-time, we have considered, are
gauge fields with a U(1) internal symmetry group. Because of the
non-commutativity of $U_{\Re }$, the expressions for gauge potential, gauge
field and action, in the case of a non-Abelian internal group, are exactly
the same. The only change is that now the coefficients $A_{a}$ and $F_{ab}$
are in $U_{\Re }\otimes L_{G}$, $L_{G}$ being the Lie algebra of the
internal symmetry group.

To discuss matter fields one also needs spinors, and an appropriate set of $%
\gamma $ matrices to contract the derivations $\partial _{a}$. A massless
action term for spinor matter fields may then be written 
\begin{equation}
S_{\psi }=i\overline{\psi }\gamma ^{a}\partial _{a}\psi  \label{8.37}
\end{equation}
where $a\in (0,1,2,3,4)$, $\gamma ^{a}=(\gamma ^{0},\gamma ^{1},\gamma
^{2},\gamma ^{3},i\gamma ^{5})$ and $\psi $ is a field in a projective
module $E_{\psi }\subset $ $U_{\Re }^{\otimes 4}$. It follows from the
properties (\ref{2.6}) of the derivations that this term is Lorentz
invariant. Notice that although the set $\{M_{\mu \nu },x_{\mu }\}$ has a
O(1,4) structure, it is only the O(1,3) part that is a symmetry group.
Coupling the fermions to the gauge fields 
\begin{equation}
S_{\psi }=\overline{\psi }i\gamma ^{a}(\partial _{a}+igA_{a}.\tau )\psi
\label{8.38}
\end{equation}
$\{\tau \}$ is a set of representatives of the internal symmetry Lie
algebra. From (\ref{8.38}) one sees that fermions may be coupled to the
connection $A_{a}$ without having to introduce new degrees of freedom in the
fermion sector.

The existence of the additional degree of freedom on the connections is a
consequence of the non-commutative space-time algebra which does not depend
on the magnitude of $\ell $ but just on $\ell $ being $\neq 0$. Therefore,
in addition to the specific effects coming from the non-commutativity of the 
$U_{\Re }$ algebra, a more dramatic consequence is the emergence of new
interactions which, for each gauge model, follow from the gauge principle.

Connes and Lott\cite{Connes1} and several other authors after them (for a
review and references see \cite{Kastler}\cite{Schucker}) have used the $%
\left( H,\pi ,D\right) $ scheme to construct a geometric formulation of the
standard model. What essentially is done is to consider as geometric space
the product of a commutative 4-dimensional space-time with a discrete space.
This is interpreted as a multi-sheeted space-time and the Yukawa coupling
matrix of the standard model provides the part of the $D$ operator that acts
on the discrete space. This construction provides a nice bookkeeping of the
Higgs phenomena and the Kobayashi-Maskawa matrix although it is a little
hard to believe that a phenomenological model, with so many free parameters
as the standard model, is a direct manifestation of the intrinsic geometry
of space-time.

The additional degrees of freedom arising from the non-commutative
space-time structure that I have been discussing have a quite different
origin from those in the Connes-Lott construction. They are not associated
to any discrete component in space-time but to the structure of the
derivations and the differential algebra. Also there is no reason to expect
them to be related to the Higgs phenomenon. In fact if the structure that we
obtained from the structural stability idea is a good clue to the behavior
of Nature, we would expect the extra degrees of freedom to manifest
themselves at the same level of dimensions as the size of the fundamental
length parameter. They are a consequence of the structure of the
differential algebra and they appear both in the abelian and non-abelian
theories, through the fields that are connections. Of particular interest
from the experimental point of view is the pseudoscalar partner of the
photon associated to the $A_{4}$ component of the Abelian connection.

The only non-trivial action of the derivation $\partial _{4}$ is on the
coordinate operators $x_{\mu }$. From (\ref{2.6}) one sees that in the $\ell
\rightarrow 0$ limit this action vanishes and all the effects of the $%
\partial _{4}$ derivation (extra connection components, etc.) disappear
together with the non commutativity of $[x_{\mu },x_{\nu }]$. Extracting the 
$\ell $ factor from $\partial _{4},\theta ^{4}$ and $A_{4}$ we estimate the
dependence on $\ell $ of all the physical effects arising from the
non-commutative space-time structure. Define 
\[
\begin{array}{c}
\overline{\partial _{4}}=\frac{1}{\ell }\partial _{4} \\ 
\overline{\theta ^{4}}=\ell \theta ^{4} \\ 
\overline{A^{4}}=\frac{1}{\ell }A^{4}
\end{array}
\]
Then 
\[
\begin{array}{c}
\overline{\partial _{4}}(x_{\mu })=P_{\mu } \\ 
\overline{\theta ^{4}}(\overline{\partial _{4}})=1 \\ 
\nabla (1)=A_{\mu }\theta ^{\mu }+\overline{A_{4}}\overline{\theta ^{4}}
\end{array}
\]
The non-abelian QED field equation 
\[
\partial _{a}F_{ab}-[A^{a},F_{ab}]=0 
\]
becomes 
\begin{equation}
\partial ^{\mu }F_{\mu \nu }+\ell ^{2}\left\{ \overline{\partial ^{4}}%
\overline{\partial _{4}}(A_{\nu })-\overline{\partial ^{4}}\partial _{\nu }(%
\overline{A_{4}})-\overline{\partial ^{4}}[\overline{A_{4}},A_{\nu }%
]\right\} -[A^{\mu },F_{\mu \nu }]-[A^{4},F_{4\nu }]=0  \label{8.43}
\end{equation}
Notice that the last two terms are also, at most, of order $\ell ^{2}$.
Therefore $\partial ^{\mu }F_{\mu \nu }=O(\ell ^{2})$ and, for example,
eventual deviations from the masslessness of the photon are expected to be
at most $O(\ell ^{2})$. Notice however that these effects have a dependence
on the energy scale of the experiment. $\partial ^{\mu }F_{\mu \nu }$ has
dimensions (length)$^{-3}$. This means that expectation values of the
operator that multiplies $\ell ^{2}$ in Eq.(\ref{8.43}) have dimension
(length)$^{-5}$. Therefore, although the deviations from $\partial ^{\mu
}F_{\mu \nu }=0$ are $O(\ell ^{2})$, they may be strongly enhanced by the
energy scale of the experiment.

For the pseudoscalar electromagnetic coupling we have 
\[
\ell g\overline{\psi }\gamma ^{5}\overline{A_{4}}\psi 
\]
that is, expected effects are of order $\ell ^{2}\alpha $. The same
considerations as above, concerning the experimental energy scale, apply to
the pseudoscalar coupling effect.

\subsection{Metric and Riemannian structure}

To complete this survey of geometric notions, a brief discussion is also
included on how to construct metrics and a Riemannian structure in
non-commutative space-time. Applications to general relativity will be
discussed elsewhere.

Besides the exterior differential algebra defined above we will be concerned
here with the tensor algebra constructed from the same $\{\theta ^{a}\}$
basis by formal tensor products 
\[
\theta ^{a_{1}}\otimes \theta ^{a_{2}}\otimes \cdots \otimes \theta ^{a_{p}} 
\]
In particular a {\it metric} is a symmetric element 
\begin{equation}
g_{kl}\theta ^{k}\otimes \theta ^{l}  \label{8.50}
\end{equation}
with $g_{kl}\in U_{\Re }$ and $g_{kl}=g_{lk}$.

We now consider $U_{\Re }$-modules constructed from elements of the tensor
algebra. To define a connection, in modules constructed in this way, it
suffices to consider the action of $\nabla _{\delta }$ in the basis elements 
$\theta ^{a}$. Define 
\begin{equation}
\nabla _{\delta }(\theta ^{a})=-\Gamma _{b}^{a}(\delta )\theta ^{b}
\label{8.51}
\end{equation}
with $\delta \in V$, $\Gamma _{b}^{a}(\delta )\in U_{\Re }$, $\theta ^{b}\in
\Omega ^{1}(U_{\Re })$. $\Gamma _{b}^{a}(\delta )$ is an element of the
enveloping algebra $U_{\Re }$ which depends on the derivation $\delta $.
Therefore it may be written as the contraction of a one-form 
\begin{equation}
\nabla (\theta ^{a})=-\Gamma _{bc}^{a}\theta ^{b}\otimes \theta ^{c}
\label{8.52}
\end{equation}
$\Gamma _{bc}^{a}\in U_{\Re }$. For a metric, using (\ref{8.24}) and the
Leibnitz rule 
\begin{equation}
\nabla (g_{bc}\theta ^{b}\otimes \theta ^{c})=\partial _{a}(g_{bc})\theta
^{a}\otimes \theta ^{b}\otimes \theta ^{c}-g_{bc}\Gamma _{de}^{b}\theta
^{d}\otimes \theta ^{e}\otimes \theta ^{c}-g_{bc}\Gamma _{de}^{c}\theta
^{d}\otimes \theta ^{b}\otimes \theta ^{e}  \label{8.53}
\end{equation}
For a metric with vanishing covariant derivative $\nabla (g_{bc}\theta
^{b}\otimes \theta ^{c})=0$ and a symmetric connection $\Gamma
_{bc}^{a}=\Gamma _{cb}^{a}$ one obtains, permuting the indices 
\begin{equation}
g_{ab}\Gamma _{de}^{a}=\frac{1}{2}\{\partial _{e}(g_{bd})+\partial
_{d}(g_{eb})-\partial _{b}(g_{de})\}  \label{8.54}
\end{equation}
which corresponds, in this setting, to the Christofell relations. The
factors on the left are however, in general, non-commuting elements of $%
U_{\Re }$, instead of c-numbers, and the derivations on the right are
computed by the rules of Eqs.(\ref{2.6}). Hence the relations are not so
useful as in the commutative case, to compute the connection from a given
metric, because the $U_{\Re }-$valued matrix $g_{ab}$ is, in general, not
easy to invert. Therefore it is more convenient to define physical
configurations from the connection itself.

The quantity that corresponds to the Riemann tensor is a $U_\Re -$valued
tensor 
\begin{equation}  \label{8.55}
R=R_{bce}^a\theta ^b\otimes \theta ^c\otimes \theta ^e\otimes \partial _a
\end{equation}
$\theta ^a\in \Omega ^1(U_\Re )$, $\partial _a\in $Der$(U_\Re )$ and $%
R_{bce}^a\in U_\Re $, which is obtained computing 
\begin{equation}  \label{8.56}
\nabla _\delta (\nabla _{\delta ^{^{\prime }}}a_b\theta ^b)-\nabla _{\delta
^{^{\prime }}}(\nabla _\delta a_b\theta ^b)-\nabla _{[\delta ,\delta
^{^{\prime }}]}a_b\theta ^b
\end{equation}
The result is 
\begin{equation}  \label{8.57}
R_{bce}^a=\partial _b(\Gamma _{ce}^a)-\partial _c(\Gamma _{be}^a)+\Gamma
_{bn}^a\Gamma _{ce}^n-\Gamma _{cn}^a\Gamma _{be}^n+if_{cb}^n\Gamma _{ne}^a
\end{equation}
For completeness the last term contains the structure constants of the
derivation algebra, which in this case vanish because Der$(U_\Re )$ is
Abelian. Notice that in Eq.(\ref{8.57}) the order of the factors, in the
quadratic terms, is not arbitrary because $\Gamma _{bc}^a\in U_\Re $.

From (\ref{8.57}), by contraction, one would obtain the non-commutative
versions of the Ricci tensor and the scalar curvature.

\section{Appendix. Quantum stochastic calculus based on ISO(1,1) and ISO(2)}

Consider first the ISO(1,1) case. In the representation (\ref{3.6g})
replacing $-i\frac{d}{d\mu }$ by $X^{0}$ and $\mu $ by $i\frac{d}{dX^{0}}$
one obtains 
\begin{equation}
\begin{array}{lllllll}
P^{0} & = & \frac{1}{2}\left( e^{i\frac{d}{dX^{0}}}-e^{-i\frac{d}{dX^{0}}%
}\right)  & = & \frac{1}{2}\left( T_{i}-T_{-i}\right)  & = & i\sin \left( 
\frac{d}{dX^{0}}\right)  \\ 
\Im  & = & \frac{1}{2}\left( e^{i\frac{d}{dX^{0}}}+e^{-i\frac{d}{dX^{0}}%
}\right)  & = & \frac{1}{2}\left( T_{i}+T_{-i}\right)  & = & \cos \left( 
\frac{d}{dX^{0}}\right) 
\end{array}
\label{A.1}
\end{equation}
Operating on holomorphic functions of $X^{0}$, $P$ and $\Im $ may be
interpreted as a finite difference and as an averaging operator in the
complex direction. Defining 
\begin{equation}
\begin{array}{lll}
\sin \left( \frac{d}{dX^{0}}\right)  & \circeq  & D_{-} \\ 
\cos \left( \frac{d}{dX^{0}}\right)  & \circeq  & D_{+}
\end{array}
\label{A.2}
\end{equation}
the commutation relations are 
\begin{equation}
\begin{array}{lll}
\left[ D_{-},X^{0}\right]  & = & D_{+} \\ 
\left[ D_{+},X^{0}\right]  & = & -D_{-} \\ 
\left[ D_{+},D_{-}\right]  & = & 0
\end{array}
\label{A.3}
\end{equation}

The first step in the construction of a quantum stochastic calculus, based
on this algebra, is the construction of the associated representation
spaces. Let 
\begin{equation}
\begin{array}{lll}
A & = & \frac{1}{\sqrt{2}}\left( X^{0}+D_{-}\right)  \\ 
A^{\dagger } & = & \frac{1}{\sqrt{2}}\left( X^{0}-D_{-}\right) 
\end{array}
\label{A.4}
\end{equation}
with algebra 
\begin{equation}
\begin{array}{lll}
\left[ A,A^{\dagger }\right]  & = & D_{+} \\ 
\left[ A,D_{+}\right]  & = & \frac{1}{2}\left( A-A^{\dagger }\right)  \\ 
\left[ A^{\dagger },D_{+}\right]  & = & \frac{1}{2}\left( A-A^{\dagger
}\right) 
\end{array}
\label{A.5}
\end{equation}
The vacuum is chosen to be the vector that is annihilated by $A$%
\begin{equation}
A\phi (X^{0})=\frac{1}{\sqrt{2}}\left( X^{0}+D_{-}\right) \phi (x^{0})=0
\label{A.6}
\end{equation}
which, in the Fourier transform, becomes 
\begin{equation}
-i\left( \frac{d}{d\omega }+\sinh \omega \right) \phi (\omega )=0
\label{A.7}
\end{equation}
yielding 
\begin{equation}
\phi (\omega )=\frac{1}{\sqrt{N_{\phi }}}e^{-\cosh \omega }  \label{A.8}
\end{equation}
with normalization factor 
\[
N_{\phi }=2K_{0}(2)=0.2277877...
\]
$K_{\nu }(z)$ being a modified Bessel function, and 
\begin{equation}
\phi (X^{0})=F^{-1}\left( \phi (\omega )\right) =\frac{1}{\sqrt{\pi K_{0}(2)}%
}K_{iX^{0}}(1)  \label{A.9}
\end{equation}

A basis is obtained by acting on $\phi $ with powers of $A^{\dagger }$

{\it Lemma A1.} The set $\left\{ A^{\dagger n}\phi :n=0,1,2,...\right\} $ is
an orthogonal set

Proof: Assume that up to order $n$ all $A^{\dagger n}\phi $ are known to be
orthogonal to all states of the form $A^{\dagger }...A^{\dagger
}D_{+}A^{\dagger }...A^{\dagger }\phi $ where $A^{\dagger }$ appears less
than $n$ times and $D_{+}$ one time. Then for $\alpha \leq n$%
\[
\left( A^{\dagger \alpha }\phi ,A^{\dagger n+1}\phi \right) =\left(
A^{\dagger }AA^{\dagger \alpha -1}\phi ,A^{\dagger n}\phi \right)
=...=\left( A^{\dagger \alpha }A\phi ,A^{\dagger n}\phi \right) =0 
\]
The base for the induction is provided by $\left( \phi ,A^{\dagger }\phi
\right) =0$ and $\left( D_{+}\phi ,A^{\dagger }\phi \right) =\frac{1}{2}%
\left( (A-A^{\dagger })\phi ,\phi \right) =0$. $\blacksquare $

The state $\phi $ may be used to define a probability distribution. An
operator $O$ in the enveloping algebra of $\left\{ X^{0},D_{+},D_{-}\right\} 
$ becomes a random variable with expectation $\left( \phi ,O\phi \right) $.
In particular, for $X^{0}$ the characteristic function is 
\begin{equation}
C_{X^{0}}(y)=\left( \phi ,e^{iyX^{0}}\phi \right) =\frac{K_{0}(2\cosh \frac{y%
}{2})}{K_{0}(2)}=\int e^{iyx}p(dx)  \label{A.10}
\end{equation}

The state $\phi $ is the analog of the harmonic oscillator ground state in
the Heisenberg algebra case. Although this is the state that will always be
used to define the probability structure, for the construction of the
stochastic process it is more convenient to use a different basis. Define 
\begin{equation}
\begin{array}{lll}
H_{+} & = & P^{0}+\Im \\ 
H_{-} & = & -P^{0}+\Im
\end{array}
\label{A.11}
\end{equation}
Then 
\begin{equation}
\begin{array}{lll}
\left[ X^{0},H_{+}\right] & = & -iH_{+} \\ 
\left[ X^{0},H_{-}\right] & = & iH_{-} \\ 
\left[ H_{+},H_{-}\right] & = & 0
\end{array}
\label{A.11a}
\end{equation}
and, in the representation (\ref{3.6g}) with $r=1$, one has the simple
action 
\begin{equation}
\begin{array}{lll}
X^{0}\psi (\mu ) & = & -i\frac{d}{d\mu }\psi (\mu ) \\ 
H_{+}\psi (\mu ) & = & e^{\mu }\psi (\mu ) \\ 
H_{-}\psi (\mu ) & = & e^{-\mu }\psi (\mu )
\end{array}
\label{A.12}
\end{equation}
with the scalar product in the space $V$ of square-integrable functions on
the hyperbola defined by 
\begin{equation}
\left( \phi ,\psi \right) _{\mu }=\int d\mu \phi ^{*}(\mu )\psi (\mu )
\label{A.13}
\end{equation}
An equally simple representation is obtained by Fourier transform, namely 
\begin{equation}
\begin{array}{lll}
X^{0}F(\lambda ) & = & i\lambda F(\lambda ) \\ 
H_{+}F(\lambda ) & = & F(\lambda +1) \\ 
H_{-}F(\lambda ) & = & F(\lambda -1)
\end{array}
\label{A.14}
\end{equation}

Let now $h=L^{2}(R_{+})$ be the Hilbert space of square-integrable functions
on the half-line $R_{+}=[0,\infty ]$. It is the variable in $R_{+}$ that
will be the continuous index $s$ labelling the convolution semigroup
generated by the probability distribution $p(dx)$ in (\ref{A.10}). It is
interpreted as the time parameter of a stochastic process, sum of
independent identically distributed random variables.

From (\ref{A.11a}) one constructs an infinite set of operators labelled by
functions in $h$ , with algebra 
\begin{equation}
\begin{array}{lll}
\left[ X^{0}(f),H_{+}(g)\right] & = & -iH_{+}(fg) \\ 
\left[ X^{0}(f),H_{-}(g)\right] & = & iH_{-}(fg) \\ 
\left[ H_{+}(f),H_{-}(g)\right] & = & 0
\end{array}
\label{A.15}
\end{equation}
$f,g\in h$.

These operators are made to act on a space that is a direct integral $H$ of
spaces of functions on the hyperbola indexed by functions on $h$, namely 
\begin{equation}
\psi (f)\circeq \oplus \int d\tau f(\tau )\psi (\mu _{\tau })  \label{A.16}
\end{equation}
with scalar product 
\begin{equation}
\left\langle \phi (f_{1}),\psi (f_{2})\right\rangle =\left(
f_{1},f_{2}\right) _{h}\left( \phi ,\psi \right) _{\mu }  \label{A.17}
\end{equation}
where, in the right hand side, the first scalar product is in $h$ and the
second in $V$. For $f,g\in h$ the action of the operators is 
\begin{equation}
\begin{array}{lll}
X^{0}(f)\psi (g) & = & -i\partial _{\mu }\psi (fg) \\ 
H_{+}(f)\psi (g) & = & \left( e^{\mu }\psi \right) (fg) \\ 
H_{-}(f)\psi (g) & = & \left( e^{-\mu }\psi \right) (fg)
\end{array}
\label{A.18}
\end{equation}
This action satisfies the commutation relations (\ref{A.15}).

Notice that other functional realizations of the commutation relations are
possible. For example, the operators may be made to act on a space of
square-integrable functionals over $h$ in the following way 
\begin{equation}
\begin{array}{lll}
X^{0}(f)F(g) & = & -iD_{f}F(g) \\ 
H_{+}(f)F(g) & = & (\overline{f},e^{g})F(g) \\ 
H_{-}(f)F(g) & = & (\overline{f},e^{-g})F(g)
\end{array}
\label{A.19}
\end{equation}
However in this case the difficulty is in finding a scalar product for which
the operators are symmetric. Therefore, here only the direct integral
construction will be used.

To define adapted processes the usual splittings are considered 
\begin{equation}
\begin{array}{lllllll}
h & = & L^{2}[0,\infty ) & = & L^{2}[0,s]\oplus L^{2}(s,\infty ) & = & 
h^{s}\oplus h^{(s} \\ 
H & = & H^{s}\otimes H^{(s} &  &  &  & 
\end{array}
\label{A.20}
\end{equation}
and an {\it adapted process }is a family $K=\left( K(s),s\geq 0\right) $ of
operators in $H$ such that for each $t$%
\begin{equation}
K(s)=K^{s}\otimes 1  \label{A.21}
\end{equation}
The basic adapted processes here are 
\begin{equation}
\begin{array}{lll}
X^{0}(s) & = & X^{0}(\chi _{[0,s]}) \\ 
H_{+}(s) & = & H_{+}(\chi _{[0,s]}) \\ 
H_{-}(s) & = & H_{-}(\chi _{[0,s]})
\end{array}
\label{A.22}
\end{equation}
$\chi _{[0,s]}$ being the indicator function of the interval $[0,s]$. Given
an elementary process $E_{s_{1}s_{2}}$, that is, a process that is constant
in the interval $[s_{1},s_{2}]$ and zero otherwise, the stochastic integral
is 
\begin{equation}
\int_{0}^{s}E_{s_{1}s_{2}}dK=E_{s_{1}s_{2}}\left\{ K(\min (s,s_{2}))-K(\min
(s,s_{1}))\right\}  \label{A.23}
\end{equation}
The stochastic integral of a general adapted process is obtained by
approximation by elementary processes and a limiting procedure. The
following two results characterize the properties of the stochastic integral
in the ISO(1,1) stochastic calculus

{\it Lemma A.2. }For adapted process $E_{0},E_{+},E_{-}$, $f,g\in h$ and $
\begin{array}{l}
\phi (f),\psi (g)\in H
\end{array}
$%
\begin{equation}
\begin{array}{l}
\left\langle \phi (f),\int_{0}^{t}\left\{
E_{0}dX^{0}+E_{+}dH_{+}+E_{-}dH_{-}\right\} \psi (g)\right\rangle = \\ 
=\int_{0}^{t}\left\langle \phi (f),-iE_{0}(s)\left( \partial _{\mu }\psi
\right) (g\chi _{ds})+E_{+}(s)\left( e^{\mu }\psi \right) (g\chi
_{ds})+E_{-}(s)\left( e^{-\mu }\psi \right) (g\chi _{ds})\right\rangle 
\end{array}
\label{A.24}
\end{equation}
where $\chi _{ds}$ denotes the indicator function of the interval $[s,s+ds]$%
. The proof follows from application of the operator action in (\ref{A.18}).
Using the properties of the scalar product in $H$, (\ref{A.24}) may be
rewritten as 
\[
-i(\phi ,\partial _{\mu }\psi )_{\mu }(f,E_{0}g)_{h}+(\phi ,e^{\mu }\psi
)_{\mu }(f,E_{+}g)_{h}+(\phi ,e^{-\mu }\psi )_{\mu }(f,E_{-}g)_{h}
\]

{\it Lemma A.3.} For adapted process $E_{0},E_{+},E_{-}$ define 
\begin{equation}
N_{i}(t)=\int_{0}^{t}\left\{
E_{0}^{(i)}dX^{0}+E_{+}^{(i)}dH_{+}+E_{-}^{(i)}dH_{-}\right\}  \label{A.25}
\end{equation}
Then, for $f,g\in h$ and $\phi (f),\psi (g)\in H$%
\begin{equation}
\begin{array}{l}
\left\langle N_{1}(t)\phi (f),N_{2}(t)\psi (g)\right\rangle = \\ 
=\int_{0}^{t}\left\langle N_{1}(t)\phi (f),\left\{ -iE_{0}^{(2)}(s)\left(
\partial _{\mu }\psi \right) (g\chi _{ds})+E_{+}^{(2)}(s)\left( e^{\mu }\psi
\right) (g\chi _{ds})+E_{-}^{(2)}(s)\left( e^{-\mu }\psi \right) (g\chi
_{ds})\right\} \right\rangle \\ 
+\int_{0}^{t}\left\langle \left\{ -iE_{0}^{(1)}(s)\left( \partial _{\mu
}\phi \right) (f\chi _{ds})+E_{+}^{(1)}(s)\left( e^{\mu }\phi \right) (f\chi
_{ds})+E_{-}^{(1)}(s)\left( e^{-\mu }\phi \right) (f\chi _{ds})\right\}
,N_{2}(t)\psi (g)\right\rangle \\ 
-\frac{i}{2}\int_{0}^{t}\left( \phi ,e^{\mu }\psi \right) _{\mu }\left\{ 
\overline{E}_{0}^{(1)}(s)\overline{f}(s)E_{+}^{(2)}(s)g(s)-\overline{E}%
_{+}^{(1)}(s)\overline{f}(s)E_{0}^{(2)}(s)g(s)\right\} ds \\ 
+\frac{i}{2}\int_{0}^{t}\left( \phi ,e^{-\mu }\psi \right) _{\mu }\left\{ 
\overline{E}_{0}^{(1)}(s)\overline{f}(s)E_{-}^{(2)}(s)g(s)-\overline{E}%
_{-}^{(1)}(s)\overline{f}(s)E_{0}^{(2)}(s)g(s)\right\} ds
\end{array}
\label{A.26}
\end{equation}

The proof follows from splitting the integral in its upper and lower
triangular regions plus the diagonal and using the commutation relations (%
\ref{A.15}) to compute the diagonal terms. The diagonal terms (the last two
terms in (\ref{A.26})) contain the Ito-type corrections for this stochastic
calculus which may be written symbolically as 
\begin{equation}
d\left( N_{1}N_{2}\right) =dN_{1}\cdot N_{2}+N_{1}\cdot dN_{2}+dN_{1}\cdot
dN_{2}  \label{A.27}
\end{equation}
with multiplication rules 
\begin{equation}
\begin{array}{lllll}
dX^{0}\cdot dH_{+} & = & -dH_{+}\cdot dX^{0} & = & -\frac{i}{2}dH_{+} \\ 
dX^{0}\cdot dH_{-} & = & -dH_{-}\cdot dX^{0} & = & \frac{i}{2}dH_{-}
\end{array}
\label{A.28}
\end{equation}
all others being identically zero.

The probability structure of the process is defined by the choice of a state
that is used to compute expectations. For this purpose one uses a direct sum
of states of type (\ref{A.8}) which are annihilated by the operator $A$ of (%
\ref{A.4}) 
\begin{equation}
\Omega =\oplus \int d\tau \phi _{\tau }=\oplus \int d\tau \frac{1}{N_{\phi }}%
e^{-\cosh \mu _{\tau }}  \label{A.29}
\end{equation}
This state belongs to the space $H$ and splits as follows 
\[
\oplus \int d\tau \chi _{[0,s]}\phi _{\tau }+\oplus \int d\tau \chi
_{[s,\infty )}\phi _{\tau }
\]
in the decomposition $H=H^{s}\otimes H^{(s}$. The triple $\left( H,\left\{
O_{\textnormal{adap.}}\right\} ,\Omega \right) $, where $\left\{ O_{\
textnormal{adap.}%
}\right\} $ denotes the set of adapted operators over $H$, is the
(non-commutative) probability space associated to ISO(1,1). For $X^{0}(f)$,
for example, the characteristic functional is 
\begin{equation}
C(f)=\left\langle \Omega ,e^{iX^{0}(f)}\Omega \right\rangle =\exp \left\{
\int d\tau \log \frac{K_{0}\left( 2\cosh \frac{f(\tau )}{2}\right) }{K_{0}(2)%
}\right\}   \label{A.30}
\end{equation}

In the ISO(2) case, to the $X,P$ and $\Im $ operators correspond finite
difference operators 
\begin{equation}
\begin{array}{lll}
\Delta _{-}f(x) & = & \frac{1}{2}\left( f(x+1)-f(x-1)\right)  \\ 
\Delta _{+}f(x) & = & \frac{1}{2}\left( f(x+1)+f(x-1)\right) 
\end{array}
\label{A.31}
\end{equation}
and the representation 
\begin{equation}
\begin{array}{lll}
X & = & x \\ 
P & = & \frac{1}{i}\Delta _{-} \\ 
\Im  & = & \Delta _{+}
\end{array}
\label{A.32}
\end{equation}
with associated operators 
\begin{equation}
\begin{array}{lll}
B & = & \frac{1}{\sqrt{2}}\left( x+\Delta _{-}\right)  \\ 
B^{\dagger } & = & \frac{1}{\sqrt{2}}\left( x-\Delta _{-}\right) 
\end{array}
\label{A.33}
\end{equation}
An alternative representation as operators acting on functions on the circle
is 
\begin{equation}
\begin{array}{lll}
X & = & i\frac{d}{d\theta } \\ 
P & = & \sin \theta  \\ 
\Im  & = & \cos \theta 
\end{array}
\label{A.34}
\end{equation}
$X$ has a discrete spectrum, with eigenvectors $e_{n}=\frac{1}{\sqrt{2\pi }}%
e^{in\theta }$, and, when this basis is used, states are in $\ell ^{2}(Z)$.
The state that is annihilated by $B$ is 
\begin{equation}
\phi _{0}=\sum_{n}c_{n}^{(0)}e_{n}  \label{A.35a}
\end{equation}
with $c_{n}^{(0)}\sim I_{n}(1)$, $I_{n}$ being the modified Bessel function.
In the representation (\ref{A.34}) 
\begin{equation}
\phi _{0}=\frac{1}{\sqrt{N_{\phi }}}e^{\cos \theta }  \label{A.35b}
\end{equation}
$N_{\phi }$ being the normalization factor $2\pi I_{0}(2)$.

As in ISO(1,1) the set $\left\{ B^{\dagger n}\phi _{0}\right\} $ is an
orthogonal set. However for the construction of the processes it is more
convenient to use arbitrary square-integrable functions on the circle and
the representation (\ref{A.34}).

$X$ is a random variable with characteristic function 
\begin{equation}
C(s)=\left\langle \phi _{0},e^{isX}\phi _{0}\right\rangle =\frac{I_{0}\left(
2\cos \frac{s}{2}\right) }{I_{0}(2)}  \label{A.36}
\end{equation}
To construct processes and a stochastic calculus, the commutation relations
are lifted to an infinite set indexed by functions on the circle

\begin{equation}
\begin{array}{lll}
\left[ X(f),V_{+}(g)\right] & = & -V_{+}(fg) \\ 
\left[ X(f),V_{-}(g)\right] & = & V_{-}(fg) \\ 
\left[ V_{+}(f),V_{-}(g)\right] & = & 0
\end{array}
\label{A.37}
\end{equation}
where $V_{+}$ and $V_{-}$ correspond to $\Delta _{+}+\Delta _{-}$ and $%
\Delta _{+}-\Delta _{-}$. These operators are made to act in a direct sum
space $H^{^{\prime }}$ of functions on the circle

\begin{equation}
\begin{array}{lll}
X(f)\psi (g) & = & i\partial _{\theta }\psi (fg) \\ 
V_{+}(f)\psi (g) & = & \left( e^{i\theta }\psi \right) (fg) \\ 
V_{-}(f)\psi (g) & = & \left( e^{-i\theta }\psi \right) (fg)
\end{array}
\label{A.38}
\end{equation}
and the construction follows the same steps as before.

\end{document}